\documentclass[pra,aps,twocolumn,superscriptaddress,showpacs,eqsecnum,nofootinbib,longbibliography]{revtex4-2}

\usepackage{cmap}
\usepackage{graphics,graphicx,epsfig}
\usepackage{epstopdf}
\usepackage[centertags]{amsmath}
\usepackage{amsfonts}
\usepackage{amssymb}
\usepackage{amsthm,color}
\usepackage{euscript}
\DeclareGraphicsExtensions{.pdf,.png,.jpg,.eps}
\usepackage[T1]{fontenc}
\usepackage[utf8]{inputenc}
\usepackage{xcolor}
\usepackage{hyperref}

\usepackage[english]{babel}

\usepackage{amsmath}
\usepackage{graphicx}
\usepackage{color}
\usepackage{hyperref}
\usepackage{ulem}

\begin{document}

\title{Excitation spectrum of a multilevel atom\\ coupled with a dielectric nanostructure}

\author{N.A. Moroz}
\affiliation{Quantum Technology Centre, M.V.~Lomonosov Moscow State University\\ Leninskiye Gory 1-35, 119991, Moscow, Russia}
\author{L.V. Gerasimov}
\affiliation{Quantum Technology Centre, M.V.~Lomonosov Moscow State University\\ Leninskiye Gory 1-35, 119991, Moscow, Russia}
\affiliation{Centre for Interdisciplinary Basic Research, HSE University, St. Petersburg 190008, Russia}
\affiliation{Centre for Advanced Studies, Peter the Great St. Petersburg Polytechnic University\\ Polytechnicheskaya 29, 195251, St. Petersburg, Russia}
\author{A.D. Manukhova}
\affiliation{Department of Optics, Palack\'{y} University, 17 Listopadu 12, 771 46 Olomouc, Czech Republic}
\author{D.V. Kupriyanov}\email{kupriyanov@quantum.msu.ru}
\affiliation{Quantum Technology Centre, M.V.~Lomonosov Moscow State University\\ Leninskiye Gory 1-35, 119991, Moscow, Russia}
\affiliation{Centre for Interdisciplinary Basic Research, HSE University, St. Petersburg 190008, Russia}
\affiliation{Department of Physics, Old Dominion University\\ 4600 Elkhorn Avenue, Norfolk, Virginia 23529, USA}

\date{\today}
\begin{abstract}
\noindent We develop a microscopic calculation scheme for the excitation spectrum of a single-electron atom localized near a dielectric nanostructure. The atom originally has an arbitrary degenerate structure of its Zeeman sublevels on its closed optical transition and we follow how the excitation spectrum would be modified by its radiative coupling with a mesoscopicaly small dielectric sample of arbitrary shape. The dielectric medium is modeled by a dense ensemble of $V$-type atoms having the same dielectric permittivity near the transition frequency of the reference atom.  Our numerical simulations predict strong coupling for some specific configurations and then suggest promising options for quantum interface and quantum information processing at the level of single photons and atoms. In particular, the strong resonance interaction between atom(s) and light, propagating through a photonic crystal waveguide, justifies as realistic the scenario of a signal light coupling with a small atomic array consisting of a few atoms. As a potential implication, the directional one-dimensional resonance scattering, expected in such systems, could provide a quantum bus by entangling distant atoms integrated into a quantum register.
\end{abstract}
\maketitle
\section{Introduction}
\noindent A robust, stable and controllable interface between quantum systems of different nature provides an important element for quantum information processing and quantum computing \cite{Kimble2008,Polzik2010,Ritsch2013}. The coupled atomic and light subsystems reveal the well-developed physical platforms for realization of various interface protocols due to their extensive research, predictable behavior, and widespread use in experiments. Even in the earlier era of quantum optics localized atoms allowed to observe intrinsically quantum phenomena, including the nonclassical statistical characteristics of light emitted by individual atoms \cite{Mandel1977} and the reversible vacuum Rabi oscillations shared by single atom and photon \cite{Klein1987}. Recently, deterministically assembled arrays of neutral atoms in optical tweezers have emerged as a promising resource for quantum simulation of many-body problems \cite{Browaeys2020} as well as to realize programmable and scalable processors for quantum computing \cite{Bluvstein2022-ic,Graham2022}.
But despite the impressive perspectives, outlined by the pioneering works, and remarkable experimental progress of last decades, the realization of an advanced quantum interface protocol, with on-demand manipulation by the quantum states of atoms and photons, remains a quite challenging experimental task.


The interaction between light and a single atom obeys the basic principles of quantum electrodynamics and is mainly observed via spontaneous emission, absorption, and scattering of a single photon by an atom. However, in free space, the efficiency of these processes is critically constrained by a small ratio of the scattering cross section to the spot area illuminating the atom by a tightly focused light beam. In general the weakness of the dimensionless coupling strength, characterizing the weakness of interaction between the atomic electron and quantized field, is basically limited by a small value of the fine-structure constant. 

The radiative coupling could be significantly enhanced and directionally controlled if the atom was placed near the dielectric nanostructure designed as a subwavelength waveguide or resonator \cite{Sukenik1992,Sukenik1993,Solano2017,Zoller2017,Lukin2020}. Such an option has encouraged experimental studies towards searching various hybrid platforms for interface protocols assisted by the interfering of atoms with mesoscopic dielectric samples. The first experiments towards subwavelength waveguide quantum electrodynamics (QED) were carried out with nanoscale silica fibers \cite{Nieddu2016,Solano2017,Klimov2002,Balykin2005,Hakuta2005}. The permittivity of silica is approximately constant within a broad spectral domain, that provides a homogeneous spatial profile of the evanescent field, effective trapping of atoms near the fiber surface, and enhanced coupling with the evanescent field. \cite{Hakuta2004,Rauschenbeutel2010,Kimble2012,Nayak2007,Nayak2008,Hakuta2012}.

However for an axially symmetric and homogeneous nanofiber the coupling with the guided mode is still weak and nanostructures with spatially inhomogeneous dielectric properties can give additional advantages. Such nanostructures enable a control over the dispersion dependence of the mode frequency as a function of its wavenumber, which, in certain conditions, should slow down the light propagation and enhance its interaction with the atoms. For nanofibers, this is technically attained by spatial modulation of the dielectric profile along the direction of light propagation, i.e. by using the properly designed photonic crystal waveguide (PCW). Stronger radiation coupling of atoms with such a one-dimensional photonic crystal via interaction with the evanescent field of its guided mode was predicted and experimentally verified in \cite{Kimble2014}. Afterwards this effect inspired a number of experimental studies towards advanced integration of atoms and nanostructures into hybrid quantum platforms preventing the external spontaneous losses and pursuing the directional light - matter interface \cite{Rauschenbeutel2021a,Rauschenbeutel2021b,Laurat2022a,Laurat2022b}. 

In the first experiments an "alligator"-type PCW was used to observe enhanced emission associated with a dielectric sample periodically and symmetrically modulated in transverse direction  \cite{Kimble2016}. However the nanostructure can be purposely designed with violated transverse symmetry and produce quite specific dispersion relations varied from the flat dependence to the cone-type profile \cite{Viktorovitch2018}. Recently the comb-type configured PCW was suggested in order to enhance the radiative coupling of trapped atoms with the  guided mode \cite{Laurat2022b}.

Here we are aiming to examine the potential for such advanced schemes of radiative coupling by the subsequent microscopic analysis. There are the following reasons why we are motivated to do that beyond the empirical argument normally applied to explain the complicated physics of atom-field interaction in the presence of nanostructures. Firstly, at present most of the theoretical approaches in a cavity and waveguide QED focus on atoms modeled either by two-level quantum systems or by a bit more general $V$-type or $\Lambda$-type energy configurations. Such transition schemes can only qualitatively reproduce the real energy structure of alkali-metal atoms, which, in turn, would be inappropriate to ignore in complicated dynamics of any realistic interface protocol. The second reason is that an accurate QED analysis, which is evidently needed for such complex quantum systems, would be difficult to compromise with empirical description, assuming a universal split of decay rate for emission into the waveguide and outer modes, and mainly supported by arguments of conventional macroscopic electrodynamics.

To overcome both these difficulties we intend to model the dielectric nanostructure by an ensemble of either randomly or regularly distributed $V$-type atoms, whose locations would be bounded by a surface having arbitrary shape and in some cases can be specifically modulated in one dimension in space (photonic crystal). Our central assumption is that there is only one fitting parameter, suggested by the model, namely the dielectric permittivity of the approached dielectric medium in a given spectral domain. In numerical simulations the appropriate dielectric permittivity could be mediated by various choices of the transition frequency of the medium atoms, their density and transition dipole moments.

The paper is organized as follows: In Section \ref{Section_II} we describe our theoretical model. In Appendix \ref{Appendix_A} we clarify how the actual dielectric medium can be microscopically approached by its artificial replica. In Appendix \ref{Appendix_B} we clarify our estimate for the van-der-Waals interaction of the atom in its ground state with a nanostructure of arbitrary shape. In Section \ref{Section_III} we present the results of our numerical simulations, which were performed for alkali metal atoms and for the nanostructures available for experimental verification. Finally we make some concluding remarks in the context of further applicability to the problems of quantum interface, light storage and quantum computations.

\section{Theoretical framework}\label{Section_II}

\subsection{The atom propagator}
\noindent The dynamics of an excited atom, treated as an open system, can be rigorously described by its propagation function (causal-type Green's function) defined as a chronologically ordered product of time-dependent $\psi$-operators averaged over the variables of external subsystems (i.e. the electromagnetic field and other matter environment)
\begin{equation}
iG_{n,n'}(\mathbf{r},t;\mathbf{r}',t')=\langle T\psi_{n}(\mathbf{r},t)\,\psi_{n'}^{\dagger}(\mathbf{r}',t')\rangle,%
\label{2.1}
\end{equation}
where both the creation $\psi_{n'}^{\dagger}(\mathbf{r}',t')$ and annihilation $\psi_{n}(\mathbf{r},t)$ operators in the Heisenberg picture are considered respectively at the atom's positions $\mathbf{r}'$ and $\mathbf{r}$ and dressed by its interaction with the quantized field and by its coupling with the medium. Here $n$ and $n'$ enumerate the internal quantum states of the atom, such that its Green's function (\ref{2.1}) is defined in the energy representation of the atom's undisturbed internal Hamiltonian. In non-relativistic description the defined causal-type Green's function coincides with the retarded type propagator, such that (\ref{2.1}) is only nonzero for $t>t'$.

Let us assume that the atom is slowed down and can be treated as an infinitely heavy and immobile particle. Then $\mathbf{r}'=\mathbf{r}$ and propagator (\ref{2.2}) reveals the fundamental solution of a Dyson (Schr\"{o}dinger-type) equation for the atom's valent electron driven by both the internal field and radiation coupling with the environment. In the energy representation this equation reads
\begin{eqnarray}
\lefteqn{G_{n,n'}(\mathbf{r},t;\mathbf{r}',t')=G_{n,n'}(\mathbf{r},t-t')\,\delta(\mathbf{r}-\mathbf{r}')}%
\nonumber\\%
&&\left[i\hbar\frac{\partial}{\partial t}-E_n\right]G_{n,n'}(\mathbf{r},t-t')%
\nonumber\\%
&&\hspace{0.5cm}-[\Sigma\ast G]_{n,n'}(\mathbf{r},t-t')=\hbar\,\delta_{n,n'}\,\delta(t-t'),%
\label{2.2}%
\end{eqnarray}
where we have assumed the time homogeneous conditions. Last term in the left-hand side contains a matrix integral operator with kernel $\Sigma_{nn''}(\mathbf{r},\tau=t-t'')$ and the asterisk denotes its convolution with the Green's function over the time argument $t''$ and matrix product over the quantum states $n''$. In the long wavelength approximation, which we will follow, this operator expresses the radiation self-action of the atom, considered as a compound system, together with its radiation coupling with other atoms belonging to the medium. The latter is intrinsically inhomogeneous so this operator depends on the atom's position.

The equation can be transformed to more informative algebraic form after its Fourier transform
\begin{equation}
G_{n,n'}(\mathbf{r},E)=\int_{-\infty}^\infty d\tau\,\mathrm{e}^{\frac{i}{\hbar}E\,\tau}\, G_{n,n'}(\mathbf{r},\tau)%
\label{2.3}%
\end{equation}
so we get
\begin{eqnarray}
\lefteqn{\left[E-E_n\right]G_{n,n'}(\mathbf{r},E)}%
\nonumber\\
&&-\sum_{n''}\Sigma_{n,n''}(\mathbf{r},E)G_{n'',n'}(\mathbf{r},E)=\hbar\,\delta_{n,n'},%
\label{2.4}%
\end{eqnarray}
where
\begin{equation}
\Sigma_{n,n''}(\mathbf{r},E)=\int_{-\infty}^\infty d\tau\,\mathrm{e}^{\frac{i}{\hbar}E\,\tau}\, \Sigma_{n,n''}(\mathbf{r},\tau).%
\label{2.5}
\end{equation}
That clarifies the physics of the Dyson equation, which can be consequently derived by the Feynman diagram method. The $\Sigma$-operator is now a matrix, visualized by a set of tight diagrams, and therefore normally referred to as a \textit{self-energy part} of the diagram sequence. If it was a fairly Hermitian operator, then equation (\ref{2.4}) would reveal us an example of a conventional Hamiltonian eigenstate problem. In such a simplification we can associate the self-energy part with correction of the atomic energy (radiation shift) induced by the spontaneous radiation process. That can be equally treated as radiative correction of the electron's energy and let us refer to this term as a \textit{mass operator}. In the considered case it corrects the original self-energy of the valence electron when the atom transforms to an exciton-type quasi-particle localized near a macroscopic or mesoscopic object. However in reality the self-energy part is non-Hermitian operator and its anti-Hermitian part is responsible for the spontaneous decay of the state.

Formally the self-energy part of a quasi-particle Green's function results from a diagram expansion following the methods of the quantum field theory extrapolated to statistical physics, see \cite{LfPtIX,LfPtX}. But in reality, it would be not so easy to do, since the external interaction with a macroscopic sample as well as the sample's internal dynamics is quite complicated and its Hamiltonian can be only schematically defined. Thus the entire description is normally accompanied by many simplifications.

We suggest a certain alternative to the approach guided by statistical physics and explain our model in the next section.

\subsection{Cooperative dynamics of the reference atom and dielectric medium}

\noindent Let us imagine an ideal dielectric medium, described by only one physical parameter, namely, by dielectric permittivity $\varepsilon$, which is assumed to be constant within the relevant spectral domain i.e. near the resonant frequency of the reference atom. Then we can approximate such a medium by a simple model approaching it by an ensemble of immobile disordered atoms having two level energy structure. For such a medium we can derive the spectrally dependent dielectric permittivity $\epsilon=\epsilon(\omega)$, as it is described in \cite{SKKH2009,KSH2017}, and fix its value near the resonance frequency of the reference atom, see Appendix \ref{Appendix_A} for details.

Since all the atoms are immobile we can consider the resolvent operator of the system Hamiltonian $\hat{H}$ as matrix acting in a subspace of finite dimension
\begin{eqnarray}
\hat{R}(E)&=&\frac{1}{E-\hat{H}}%
\nonumber\\%
\tilde{\hat{R}}(E)&=&\hat{P}\,\hat{R}(E)\,\hat{P},%
\label{2.6}%
\end{eqnarray}
where the second line projects the global resolvent operator $\hat{R}(E)$ onto the linear span, sharing the single excitation within the atomic subsystem, and onto the vacuum field state $|0\rangle_{\mathrm{Field}}$. The projector $\hat{P}$ can be decomposed in two terms
\begin{equation}
\hat{P}=\left(\hat{P}_A+\hat{P}^{(N)}\right)|0\rangle\langle 0|_{\mathrm{Field}},%
\label{2.7}%
\end{equation}
where $\hat{P}_A$ projects on an excited state of the reference atom
\begin{equation}
\hat{P}_A=\sum_{n}|n;g\ldots g\rangle\langle g\ldots g;n|%
\label{2.8}%
\end{equation}
when all other $N$ atoms, belonging to the medium, occupy the collective ground state $|g\ldots g\rangle$ and
\begin{equation}
\hat{P}^{(N)}=\sum_{a=1}^{N}\sum_{m,e}|m;g\ldots {e}\big{|}_a\ldots g\rangle\langle g\ldots {e}\big{|}_a\ldots g;m|%
\label{2.9}%
\end{equation}
projects on any ground Zeeman state $|m\rangle$ of the reference atom and each, but only one, atom of the medium can occupy the excited state $|e\rangle$, such that we have subscribed it in (\ref{2.9}) by the atom's position $a$ running from $1$ to $N$.

In the cooperative dynamics of the reference atom and the medium the excitation virtually migrates among all the atoms and the resolvent operator $\tilde{\hat{R}}(E)$ can be expressed by the Fourier image of the multi-particle Green's function. The latter is defined as the following time ordered product of the second quantized operators
\begin{eqnarray}
\lefteqn{iG^{(N+1)}(x_1,\ldots,x_N;x|x';x_N',\ldots,x_1')}%
\nonumber\\%
&&=\left\langle T\,\psi(x_1)\ldots\psi(x_N)\psi(x)\psi^{\dagger}(x')\psi^{\dagger}(x_N')\ldots\psi^{\dagger}(x_1')\right\rangle,%
\nonumber\\%
\label{2.10}
\end{eqnarray}
where we have incorporated all the quantum state specifications, including position and time arguments into the symbolic arguments $x$ for the reference atom and $x_a$ for any $a$-th atom of the medium.

Denote any accessible internal states of the reference and medium atoms as $|\alpha\rangle\equiv |m\rangle,|n\rangle,\ldots$ and $|\beta\rangle\equiv |g\rangle,|e\rangle,\ldots$ respectively. Then we can express an arbitrary matrix element of the resolvent by the following Laplace-type integral transform of the multi-particle Green's function
\begin{eqnarray}
\lefteqn{\langle\beta_1\ldots\beta_N;\alpha|\tilde{\hat{R}}(E)|\alpha';\beta_N'\ldots\beta_1'\rangle}%
\nonumber\\%
&&\hspace{1cm}\times\,\delta(\mathbf{r}-\mathbf{r}')\,\delta(\mathbf{r}_1-\mathbf{r}_1')\ldots\delta(\mathbf{r}_N-\mathbf{r}_N')%
\nonumber\\%
\nonumber\\%
&&\hspace{1cm}=\frac{1}{\hbar}\int_0^\infty\!dt\,\exp\left[+\frac{i}{\hbar}Et\right]%
\nonumber\\%
&&\hspace{-0.3cm}\times\!\left.G^{(N+1)}(x_1,\ldots,x_N;x|x';x_N',\ldots,x_1')\right|_{\scriptsize{\begin{array}{l}t_1\!=\!\ldots\!=\!t_N\!=\!t\\ t_1'\!=\!\ldots\!=\!t_N'\!=\!t'\!=\!0\end{array}}}%
\nonumber\\%
&&\label{2.11}%
\end{eqnarray}
which coincides with its Fourier transform since the integrand vanishes at $t<0$.

Now we can clarify our calculation model. We associate the single particle Green's function of the reference atom, earlier defined by (\ref{2.1})-(\ref{2.3}), with the following matrix projection of the entire resolvent
\begin{equation}
\hat{G}(\mathbf{r},E)=\hbar\,\hat{P}_A\,\tilde{\hat{R}}(E)\,\hat{P}_A,%
\label{2.12}%
\end{equation}
where the right-hand side is a microscopically defined quantity, which depends on either positions of the reference as well as medium atoms. However we expect and further verify it by our numerical simulations that, being averaged over different spatial configurations, the Green's function, defined by (\ref{2.12}), remains sensitive to specific choice of the density of medium atoms, frequency offset $\omega_{eg}-\omega_{nm}$, and internal interaction parameters only via their contribution to the self-consistently constructed dielectric permittivity of the medium $\epsilon(\omega)$ taken near vicinity of the transition frequency $\omega_{nm}$.

Considering equation (\ref{2.4}) in its matrix form we can select the self-energy part, dressed by interaction with the medium, by the following identity
\begin{equation}
\hat{\Sigma}(\mathbf{r},E)=E-E_n-\hbar\hat{G}^{-1}(\mathbf{r},E).%
\label{2.13}%
\end{equation}
For applications, it is most important to evaluate this matrix function near the resonance point $E\sim E_n$, where we obtain
\begin{equation}
\hat{\Sigma}(\mathbf{r})\equiv\hat{\Sigma}(\mathbf{r},E_n)=-\hbar\hat{G}^{-1}(\mathbf{r},E_n),%
\label{2.14}%
\end{equation}
which gives us the radiation correction to the energy structure of the reference atom localized near the medium surface at the distances having an order of the radiation wavelength or less. For the sake of notation convenience, we fix the ground state energy by zero value and denote the reference transition frequency, same for all the Zeeman states, as $\omega_{nm}\equiv\omega_0$. Then the transition frequency of a medium atom can be expressed as $\omega_{eg}\equiv\omega_M\equiv\delta_{M}+\omega_0$, where $\delta_M$ is its frequency offset in respect to the reference transition frequency. 

\subsection{The resolvent operator}

\noindent The closed equation for the resolvent operator $\tilde{\hat{R}}(E)$ can be constructed after decoding the following Dyson-type diagram equation for the collective multi-particle Green's function $G^{(N+1)}(\ldots)$:
\begin{equation}
\scalebox{0.6}{\includegraphics*{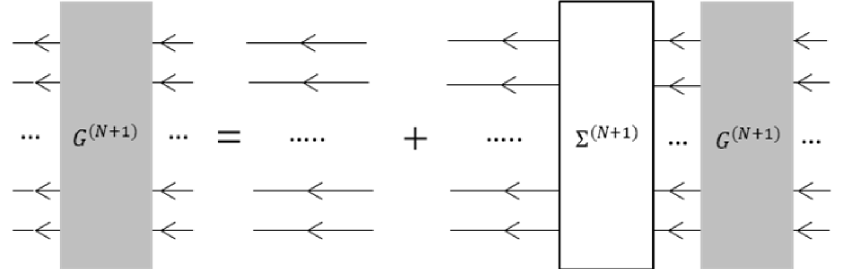}}%
\label{2.15}%
\end{equation}
Here the shaded block imagines the function $G^{(N+1)}(\ldots)$ dressed by the radiation coupling. The arrowed straight lines imagine the free propagators of either reference or medium atoms. The expansion of (\ref{2.15}) generates a sequence of virtual radiation processes transporting the single excitation within the global atomic chain. So only one of $N+1$ inward and outward arrowed lines on the shaded block belongs to an excited atom.

The key element of the Dyson equation (\ref{2.15}) is the cooperative multi-particle self-energy operator, imagined by the white block and expressed by irreducible interaction diagrams. Due to the exponential structure of the undressed individual atom propagators, the undressed many particle Green's function has exponential form as well and equation (\ref{2.15}) can be straightforwardly converted to the matrix-type operator equation defined in the unitary subspace spanning over all the atoms. Eventually for the resolvent $\tilde{\hat{R}}(E)$ we arrive at the following matrix equation
\begin{equation}
\left[(E-\hbar\omega_0)\hat{P}_A+(E-\hbar\omega_{M})\hat{P}^{(N)}-\hat{\Sigma}^{(N+1)}\right]\tilde{\hat{R}}(E)=\hat{I},%
\label{2.16}%
\end{equation}
which we intend to evaluate near the pole point associated with the atomic resonance $E\sim\hbar\omega_0$. The cooperative self-energy part $\hat{\Sigma}^{(N+1)}$ is compiled from the single-particle and two-particle basic diagrams. Below we clarify these building diagram blocks and respective partial contributions to $\hat{\Sigma}^{(N+1)}$ and enumerate them by superscribe indices running from $0$ (reference atom) to $a,b=1\ldots N$ (medium atoms).

There is a fundamental vacuum contribution to the mass operator of the reference atom
\begin{equation}
\scalebox{0.4}{\includegraphics*{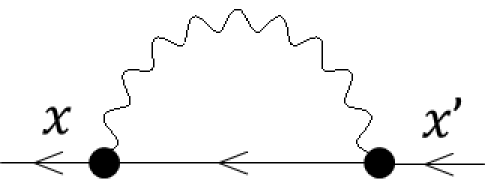}}%
\label{2.17}%
\end{equation}
which expresses the self-action on the atom by its own radiation spontaneously emitted or just virtually created in free space. The estimate of this diagram, within framework of the used long-wavelength dipole gauge, is quite nontrivial, see for example \cite{BerstLifshPitvskIV}, and eventually leads to the following result
\begin{eqnarray}
\Sigma^{(0)}_{nn'}(E)&=&\Sigma^{(0)}_{n}(E)\,\delta_{nn'}%
\nonumber\\%
\Sigma^{(0)}_{n}(E)&\approx&\Sigma^{(0)}_{n}(\hbar\omega_0)\equiv\hbar\Delta_{n}-i\hbar\frac{\Gamma_n}{2},%
\label{2.18}%
\end{eqnarray}
where
\begin{eqnarray}
\Delta_{n}&=&-\frac{2}{3\pi}\sum_{\alpha}\frac{\omega_{n\alpha}^3}{\hbar c^3}\,|\mathbf{d}_{n\alpha}|^{2}\;\ln \frac{c\,k_{\mathrm{max}}}{|\omega_{n\alpha}|}%
\nonumber\\%
\Gamma_{n}&=&\frac{4}{3}\sum_{\alpha<n}\frac{\omega_{n\alpha}^3}{\hbar c^3}\,|\mathbf{d}_{n\alpha}|^{2}=\frac{4}{3}\sum_{m}\frac{\omega_{0}^3}{\hbar c^3}\,|\mathbf{d}_{nm}|^{2},%
\nonumber\\%
\label{2.19}
\end{eqnarray}
where $\mathbf{d}_{n\alpha}$ are the transition matrix elements of the atomic dipole moment. The first line determines the logarithmic diverging energy shift where the contributing field spectrum is bounded by a cutoff wave-number $k_{\mathrm{max}}$. The cutoff $k_{\mathrm{max}}$ fulfills the inequality  $|\omega_{n\alpha}|/c\ll k_{\mathrm{max}}\lesssim a_0^{-1}\ll k_C =m_ec/\hbar$, where $a_0$ is Bohr radius and $m_e$ is electronic mass, i.e. it has to be an order of atomic scale $a_0^{-1}$ but much less than the Compton scale for a linear momentum of the electron. The sum in the expression for $\Delta_n$ expands over all the atomic levels (with including continuous part of the atomic spectrum) such that the partial contribution can be either negative if $\omega_{n\alpha}>0$ or positive if $\omega_{n\alpha}<0$. The convergence of the sum is guaranteed by the transition matrix elements vanishing for highly excited atomic states $|\alpha\rangle$. Let us stress that a similar estimate can be done for the ground state $|m\rangle$ of the reference atom either and will lead to its radiation shift. Nevertheless both the shifts have to be incorporated into physical energies of the atomic levels and in our practical calculations we leave in (\ref{2.18}) only the decay rate of the excited state $\Gamma_n$.

The elementary interaction of the reference atom with any $a$-th atom of the medium is imaged by the following diagrams
\begin{equation}
\scalebox{0.35}{\includegraphics*{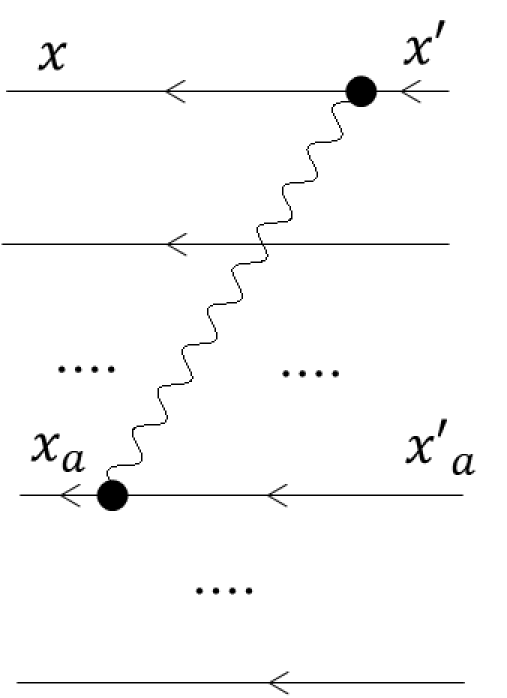}}%
\label{2.20}%
\end{equation}
and
\begin{equation}
\scalebox{0.35}{\includegraphics*{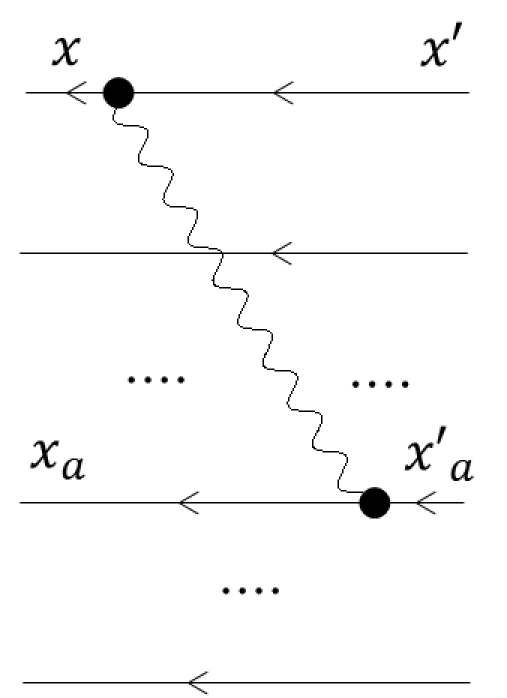}}%
\label{2.21}%
\end{equation}
where we have split one virtual process in two differently time-ordered parts.

In these diagrams the internal wavy line expresses the causal-type vacuum Green's function of the chronologically ordered polarization components of the electric field operators
\begin{equation}
iD^{(E)}_{\mu\nu}(\mathbf{R},\tau)=\left\langle T\hat{E}_{\mu}(\mathbf{r},t)%
\hat{E}_{\nu}(\mathbf{r}',t')\right\rangle,%
\label{2.22}%
\end{equation}
which depends only on difference of its arguments $\mathbf{R}=\mathbf{r}-\mathbf{r}'$ and $\tau=t-t'$. For coupling of two distant atoms it is expressed by the vacuum photon propagator with zero scalar part and with vector indices running $\mu,\nu=x,y,z$.\footnote{The field quantization is conventionally introduced in the Coulomb gauge, but by making use of the gauge invariance in the internal parts of the Feynman diagrams the field propagators can be linked with the fundamental solution of the Maxwell equation responding on a point-like dipole current, see \cite{BerstLifshPitvskIV}. This gauge has zero scalar part of the photon's propagator and is equivalent to the long-wavelength dipole gauge.} Its Fourier image is given by
\begin{eqnarray}
\lefteqn{D^{(E)}_{\mu\nu}(\mathbf{R},\omega)=\int_{-\infty}^{\infty} d\tau\,\mathrm{e}^{i\omega\tau}%
D^{(E)}_{\mu\nu}(\mathbf{R},\tau)}%
\nonumber\\%
&=&-\hbar\frac{|\omega|^3}{c^3}\left\{i\frac{2}{3}h^{(1)}_0\left(\frac{|\omega|}{c}R\right)\delta_{\mu\nu}\right.%
\nonumber\\%
&&\left.+\left[\frac{X_{\mu}X_{\nu}}{R^2}-\frac{1}{3}\delta_{\mu\nu}\right]%
ih^{(1)}_2\left(\frac{|\omega|}{c}R\right)\right\}.%
\label{2.23}%
\end{eqnarray}
Here $h^{(1)}_L(\ldots)$ with $L=0,2$ are the spherical Hankel functions of the first kind. It is important that in the self-action diagram (\ref{2.17}) the Green's function (\ref{2.22}) has contributed to the fundamental mass operator with spatial argument $\mathbf{R}\to\mathbf{0}$. The direct implication of the long-wavelength approximation would be insufficient in that specific case and, as we have pointed above, the additional physical arguments are needed for correct evaluation of diagram (\ref{2.17}).

Let us assume that in the diagrams (\ref{2.20}) and (\ref{2.21}) the reference and medium atoms are originally in the excited and ground states respectively. Then the retarded-type virtual transfer of the excitation (\ref{2.20}) leads to the following contribution to the cooperative self-energy part
\begin{eqnarray}
\Sigma^{(0a,+)}_{me;n'g}(E)&=&\int_{-\infty}^{+\infty}\frac{d\omega}{2\pi} f^{\mu}_{eg}d^{\nu}_{mn'}%
iD^{(E)}_{\mu\nu}(\mathbf{R}_{0a},\omega)%
\nonumber\\%
&&\hspace{-0.5cm}\times\frac{1}{E-\hbar\omega-E_g-E_{m}+i0}%
\label{2.24}
\end{eqnarray}
and its advanced-type counterpart (\ref{2.21}) reads
\begin{eqnarray}
\Sigma^{(0a,-)}_{me;n'g}(E)&=&\int_{-\infty}^{+\infty}\frac{d\omega}{2\pi} f^{\mu}_{eg}d^{\nu}_{mn'}%
iD^{(E)}_{\mu\nu}(\mathbf{R}_{0a},\omega)%
\nonumber\\%
&&\frac{1}{E+\hbar\omega-E_e-E_{n}+i0},%
\label{2.25}
\end{eqnarray}
where atoms are separated by a distance $R_{0a}=|\mathbf{r}_a-\mathbf{r}_0|$. Further we rename $\mathbf{r}_0\equiv\mathbf{r}$, with having in mind elimination of the medium microscopic structure in conversion (\ref{2.12})-(\ref{2.14}). For a sake of generality we use covariant notation for the tensor indices and assume the invariant sum over repeated indices. The vector components of the dipole matrix elements $d^{\nu}_{mn'}$ and $f^{\mu}_{eg}$ are related with the reference and medium atoms respectively. The complete contribution is given by sum of both the terms
\begin{equation}
\Sigma^{(0a)}_{me;n'g}(E)=\Sigma^{(0a,+)}_{me;n'g}(E)+\Sigma^{(0a,-)}_{me;n'g}(E)%
\label{2.26}%
\end{equation}
and has to be evaluated near the point $E\approx E_n=\hbar\omega_0$

Let us select the dominant contribution with constructing a delta-function singular feature in the integrand and then estimate the resting term
\begin{eqnarray}
\Sigma^{(0a)}_{me;n'g}(E)&\approx&\Sigma^{(0a)}_{me;n'g}(\hbar\omega_0)=\frac{1}{\hbar}\,f^{\mu}_{eg}d^{\nu}_{mn'}D^{(E)}_{\mu\nu}(\mathbf{R}_{0a},\omega_0)%
\nonumber\\%
&& +\ \ldots,%
\label{2.27}%
\end{eqnarray}
where ellipses denote the rest
\begin{eqnarray}
\lefteqn{\hspace{-0.5cm}\ldots=\frac{1}{\hbar}\int_{-\infty}^{+\infty}\frac{d\omega}{2\pi}\, f^{\mu}_{eg}d^{\nu}_{mn'}\,iD^{(E)}_{\mu\nu}(\mathbf{R}_{0a},\omega)}%
\nonumber\\%
&&\times\,\frac{\delta_M}{(\omega-\omega_{M}+i0)(\omega-\omega_0+i0)}.%
\label{2.28}%
\end{eqnarray}
In accordance with general constraints of the rotating wave approximation this integral expansion is representative for the frequency argument nearby the transition frequency and only if the offset $\delta_M$ is sufficiently small i.e. within assumptions that $\omega\sim\omega_0,\omega_M$ and $\delta_M\ll\omega_0,\omega_M$. Then in the spectral domain, adjacent to $\omega_0,\,\omega_M$, the integral becomes asymptotically converging and can be fairly estimated by omitting modulus in (\ref{2.23}) and substituting here the retarded-type Green's function of the field operators. That approximates integrand as analytic function in the upper half-plane of the complex $\omega$, such that (\ref{2.28}) vanishes after making a closed contour integration in the upper half-plane. So within the made approximations the resting term in (\ref{2.27}) makes only negligible correction to the main result, expressed by the first line.\footnote{\noindent This estimate would be invalid for the atoms separated by a long distance $\gtrsim c/\delta_M$. But such long distances are non attainable for our numerical simulations further performed for a sample of much less spatial scale Our evaluation of the diagrams (\ref{2.20}) and (\ref{2.21}) is also constrained by separation of the reference atom from the dielectric surface with a distance comparable with $c/\omega_0$ but not significantly shorter of it. Otherwise it should be extended by a short range chemical interaction with including the complete excitation spectrum of the medium.}

If in the diagrams (\ref{2.20}) and (\ref{2.21}) the medium atom was assumed as originally excited the similar transformations would lead to
\begin{eqnarray}
\Sigma^{(a0)}_{gn;e'm'}(E)&\approx&\Sigma^{(a0)}_{gn;e'm'}(\hbar\omega_0)%
\nonumber\\%
&=&\frac{1}{\hbar}\,d^{\mu}_{nm'}f^{\nu}_{ge'}D^{(E)}_{\mu\nu}(\mathbf{R}_{a0},\omega_0) +\ \ldots,%
\nonumber\\%
\label{2.29}%
\end{eqnarray}
where unlike of the previous case the diagram (\ref{2.21}) describes the retarded dynamics and the diagram (\ref{2.20}) has advanced status. The spatial vector argument is now directed from atom "$a$" to atom "$0$".

Similar diagrams visualize coupling between any atoms of the medium. We can define the following partial contribution between any atoms $a$ and $b$
\begin{eqnarray}
\Sigma^{(ab)}_{ge;e'g}(E)&\approx&\Sigma^{(ab)}_{ge;e'g}(\hbar\omega_0)%
\nonumber\\%
&=&\frac{1}{\hbar}\,f^{\mu}_{eg}f^{\nu}_{ge'}D^{(E)}_{\mu\nu}(\mathbf{R}_{ab},\omega_0) +\ \ldots,%
\nonumber\\%
\label{2.30}%
\end{eqnarray}
where indices $a$ and $b$ are independently running from $1$ to $N$ and each selected pair of the atoms contributes twice because of two alternative options in sharing a single excitation in the system of two atoms.

The following remarks concerning the presented derivation seem important. For the medium atoms the self-action diagram (\ref{2.17}) should be also incorporated into the entire construction of the complete self-energy operator $\hat{\Sigma}^{(N+1)}$. But in fact in far off-resonance conditions with $\delta_M\gg\Gamma_e$, where $\Gamma_e$ is the natural decay rate for a medium atom, it makes only negligible correction to evaluation of the resolvent operator. The crucial assumption, approving the domination of the interaction diagrams (\ref{2.20}) and (\ref{2.21}) with avoiding more complex graphs, is that the system dynamics would be approached as free within a short retardation time for light propagation between the atoms of any pair.  Finally we can point out that the complete self-energy operator consists of the Hermitian and anti-Hermitian parts. The former is responsible for the correction of energy structure in the multi-atomic system sharing the single excitation and the latter for the decay process of such an unstable exciton-type quasi-particle state.

\section{Results}\label{Section_III}
\noindent Here, we present two examples demonstrating capabilities of the method in application to some feasible and actively studied experimental objects. We compare our results with the alternative and independent calculations and arguments conventionally utilizing the approach of macroscopic electrodynamics.

\subsection{Cylindrical nanoscale waveguide}

\begin{figure}[pt]
\includegraphics[width=8cm]{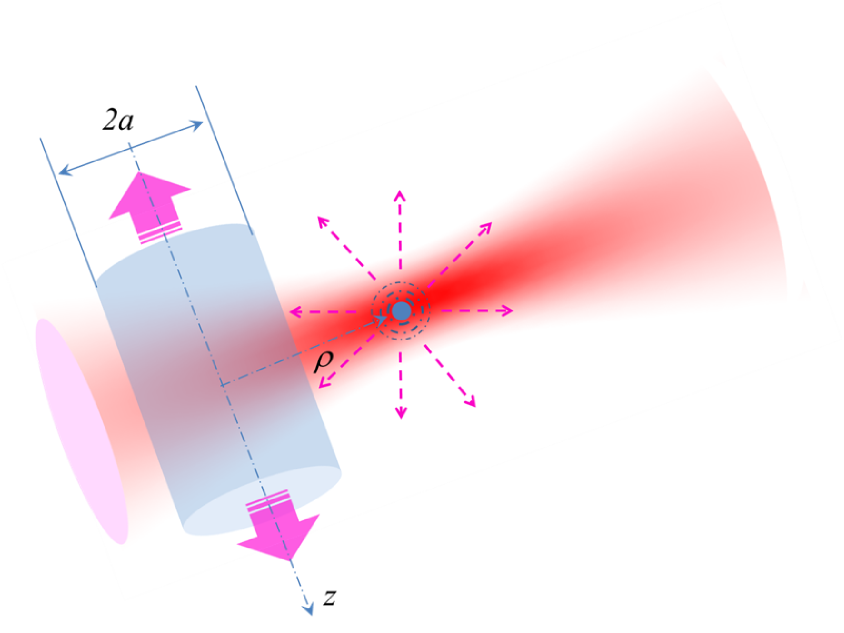}%
\caption{The single atom placed by optical tweezers near a cylindrical nanoscale waveguide. The atomic emission is partly enhanced into the guided mode.}
\label{fig1}%
\end{figure}

\noindent The mesoscopic systems, consisting of a few atoms with individual access to each of them, make convenient experimental resource for studying various QED effects that has been highlighted by recent experiments with dielectric nanostructures \cite{Beguin2020,Rauschenbeutel2021b,Rauschenbeutel2020,Laurat2022b} and microcavities \cite{Volz2020,Lukin2020}. As a first example, shown in Fig.~\ref{fig1}, we consider the optical coupling of an alkali-metal atom with a cylindrical dielectric waveguide, made of silica (SiO${}_2$ with refractive index $\mathrm{n}=1.45$). In experiments the trapping of atoms is normally designed by two beams propagating along the fiber and oppositely detuned from the atomic transition \cite{CGCGSKL2016,SBKISMPA2016}. But here we have in mind and examine a different trapping technique, based on optical tweezers, when the atom is loaded into a potential well appeared near the caustic waist of an auxiliary light beam and configured by its far off-resonant red detuned Gaussian mode. The trapping light affects and can significantly distort the atomic energy structure, which can be important for its implication as a logic unit in quantum interface or quantum computing, see \cite{Lukin2020,Gerasimov2021}. Nevertheless we further neglect the action of the trapping light on the atom and focus on a net effect i.e. on the energy correction induced by its own radiation only.

\subsubsection{Tripod configuration, ${}^{87}Rb$}

\noindent In the hyperfine manifold of $D_2$-line of ${}^{87}$Rb there is a unique configuration of the closed tripod-type transition existing between the $F=0$ excited state and $F_0=1$ ground state. \footnote{Here and throughout we specify the atomic states by total (orbital+electronic spin+nuclear spin) angular momentum and its projection $F_0,M_0$ and $F,M$ respectively for the ground and excited states.} That gives a convenient example of emission channel letting comparative analysis for alternative calculations based on the macroscopic approach. The atom has its non-degenerate upper state, so in macroscopic description we can avoid the subtle problem with evaluation of the energy shift, induced by interaction with the medium, via renormalization of the "dressed" and formally diverging upper state energy to its experimentally measured physical value. The microscopic approach, within the made approximations, leaves finite all the calculated parameters.  

The self-energy operator (\ref{2.14}) reduces to the energy correction of the the $F=0$ excited state
\begin{equation}
\Sigma(\rho)=\Delta(\rho)-\frac{i}{2}\Gamma(\rho),%
\label{3.1}%
\end{equation}
where $\Delta=\Delta(\rho)$ and $\Gamma=\Gamma(\rho)$ are the energy shift and decay rate, considered as function of radial distance $\rho$.  We have omitted here and in further examples the vacuum contribution to the shift by incorporating it into the physical energy of the free atom. Then the extra energy shift includes the radiative correction, associated with the emitted light, and estimates the near field static interaction with the sample. The asymptotic value $\Gamma(\infty)\equiv\Gamma_{\infty}$ in (\ref{3.1}) defines the decay rate of free atom.

\begin{figure}[pt]
\includegraphics[width=8cm]{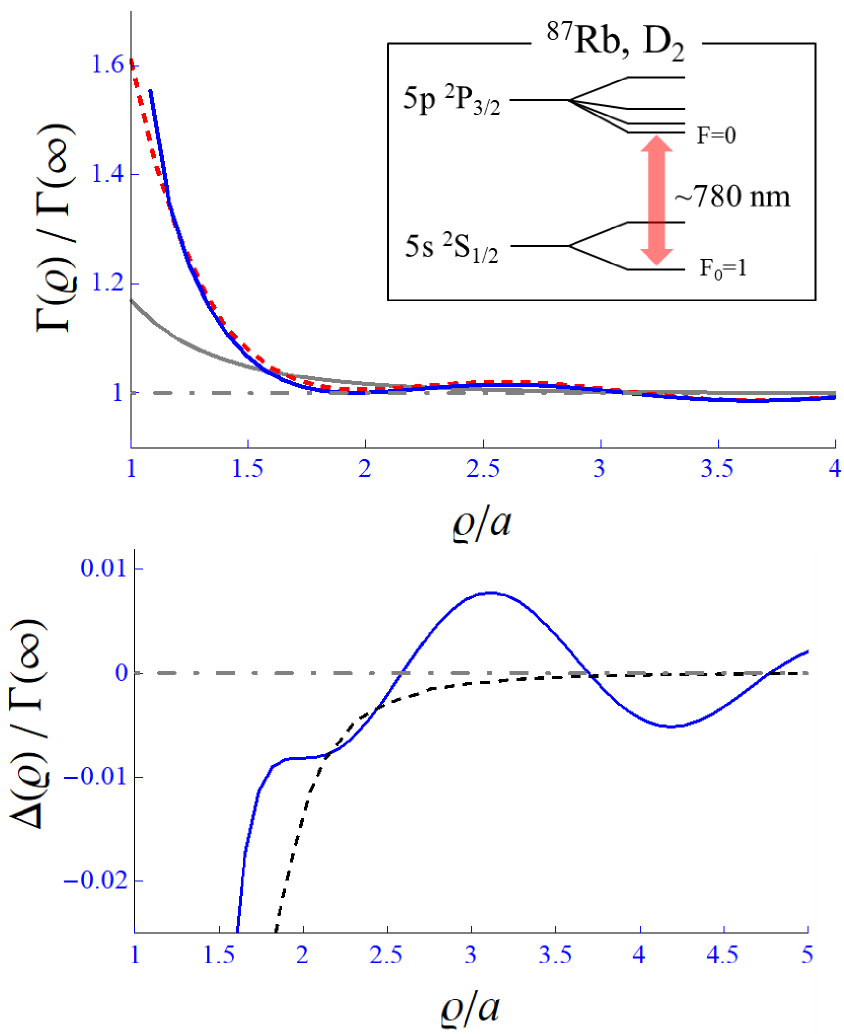}%
\caption{Upper plot: decay rate of the hyperfine sublevel $F=0$ belonging to the excited state of ${}^{87}$Rb ($D_2$-line, $\lambda\sim 780\, nm$), calculated microscopically (blue solid), by the Fermi's golden rule (red-dashed), and its  asymptotic estimate from \cite{PSGPCLK2018} (gray), see text. Lower plot: the radiative shift of the same sublevel (blue solid) and the van-der-Waals shift of the ground state (dashed). The dependencies on radial distance $\rho$ were calculated for a silica cylindrical nanofiber with radius $a=200\, nm$ (see Fig.~\ref{fig1}) and scaled by the natural decay rate $\Gamma(\infty)\equiv\Gamma_{\infty}$, approached in the limit $\rho\to\infty$.}
\label{fig2}%
\end{figure}

In Fig.~\ref{fig2} we show the results of the microscopic and macroscopic calculations for the decay rate of the $F=0$ state (upper plot) and its energy shift, microscopically calculated, (lower plot). The atom is assumed to be placed near a nanoscale waveguide (nanofiber) with radius $a=200\,nm$. In addition in the lower plot we have shown our estimate for the van-der-Waals shift of the ground state, see Appendix \ref{Appendix_B}. The actual shift of the transition frequency is given by difference of these two quantities.

The results of macroscopic calculations are adopted from \cite{PSGPCLK2018} and, in turn, were made in two different ways. In the macroscopic vision of the problem it is most naturally to apply the Fermi's golden rule and evaluate numerically the transition rate into all the set the field modes distorted by the presence of the nanofiber. These modes can be constructed by straightforward solution of the Maxwell equations, see \cite{Marcuse82}, and the final result is reproduced in the upper plot of Fig.~\ref{fig2} by the red-dashed curve. As a second way, it is possible to analytically construct a long distant asymptotic approximation to the exact result in assumptions of a certain balance in the coupling of atomic emitter with the $HE_{11}$-guided and external modes, see \cite{PSGPCLK2018}. These calculated data are reproduced by the gray curve in the plot. 

The microscopic calculations for the decay rate and light shift are given by the blue curves in both the plots of Fig.~\ref{fig2}. For the decay rate we have obtained excellent agreement between the data calculated in the microscopic and macroscopic approaches. Let us stress that the presented calculations were independently done without any fitting manipulations. In the microscopic case we have verified the internal conversion of the calculation process and its eventual insensitivity to variation of the external parameters of the artificial medium, such as atomic density, sample length and frequency offset, see Appendix \ref{Appendix_A}.  The data sets show the oscillation behavior indicating the interference of the emitted radiation and incorporation of the reference atom into a local nano-cavity system with the waveguide. But the primitive balance description of light emission, expressed by the gray curve, loses this effect. 

The main advantage of the microscopic approach is that it lets us estimate the shift of the energy level, induced by the radiative coupling of the atom with the nanofiber. At short distances this shift is masked by the van-der-Waals static attraction to the sample in both the excited and ground states. Our estimates indicate the importance of the static interaction for distances of $200\, nm$ or shorter. The net radiative correction to the energy shift is quite small within $\sim 0.01\Gamma_{\infty}$. But even such a relatively small shift can be visible in observation of microwave-optical double resonances driven by coherent fields.

\subsubsection{The $F=5\to F_0=4$ closed transition in ${}^{133}$Cs}\label{Section_IIIA2}

\noindent Consider cesium atom, coupled with the same cylindrical nanofiber. The critical difference with the previous example is that the closed emission transition $F=5\to F_0=4$ in ${}^{133}$Cs has degenerate both the upper and lower states. The interaction with the nanofiber violates the spherical symmetry, such that the excited state splits in six quasi-energy sublevels. The self-energy operator (\ref{2.14}) reveals $11\times 11$ matrix having six different eigenvalues
\begin{equation}
\Sigma_{n}(\rho)=\Delta_{n}(\rho)-\frac{i}{2}\Gamma_{n}(\rho).%
\label{3.2}%
\end{equation}
where one specific eigenvalue, labeled by $n=0$ is non-degenerate, and other five with $n=1,\ldots 5$ are double degenerate. Asymptotically in the limit of weak coupling the respective eigenstates correlate with the atomic state $|F,M=0\rangle$ and with the pair of states $|F,M\rangle$ with $|M|=1,\ldots 5$, where the relevant quantization direction is clarified below. But before and for a sake of comparative discussion let us draw attention to the earlier calculations in \cite{Balykin2005}, based on the Fermi's golden rule, where the initial and final states were intuitively defined as undisturbed atomic states with angular momentum and its projection, quantized along the waveguide direction. 

\begin{figure}[pt]
\includegraphics[width=8cm]{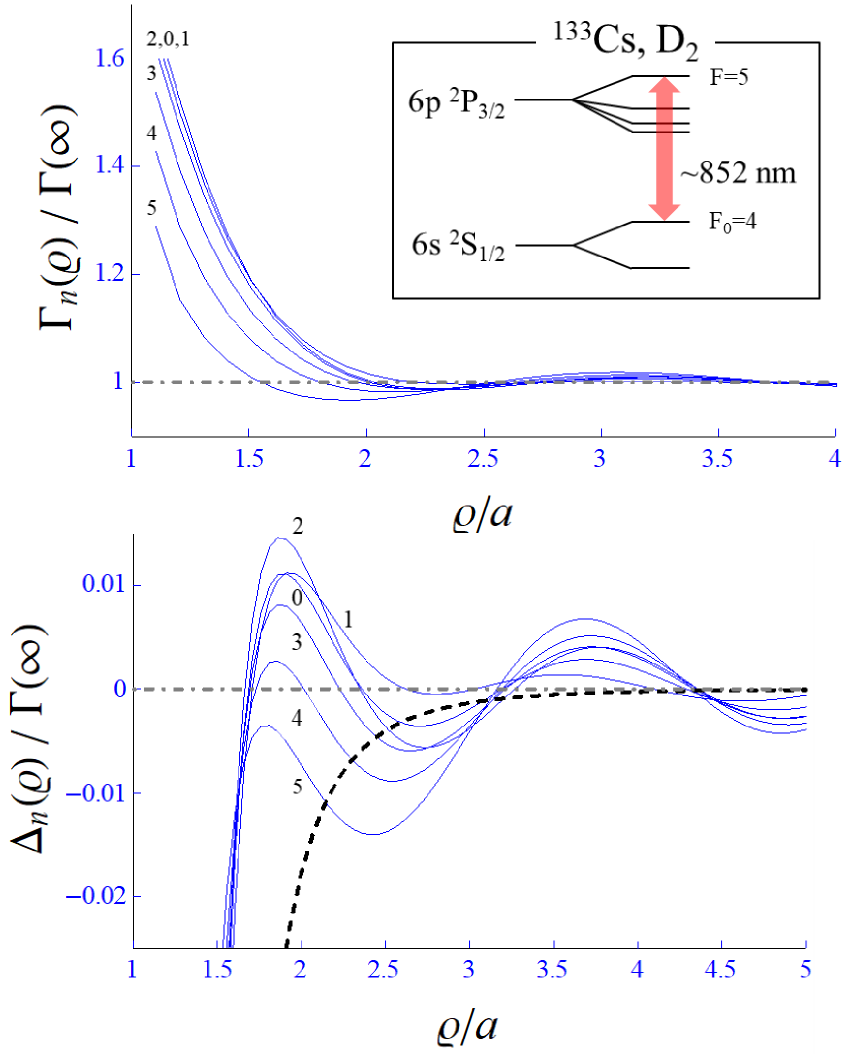}%
\caption{Same as in Fig~\ref{fig2}, but calculated for the $F=5\to F_0=4$ closed transition in $D_2$-line of ${}^{133}$Cs. The six different curves, enumerated by $n=0,1\ldots 5$, reproduce the six imaginary (upper plot) and real (lower plot) parts of the eigenvalues of the self-energy operator, i.e. the state decay rates and the radiative shifts.}
\label{fig3}%
\end{figure}

In Fig.~\ref{fig3} we present the results of our microscopic calculations for the eigenvalues (\ref{3.2}). For six different eigenvalues the obtained degeneracy results from the symmetry of emission process and the correct eigenfunctions of the operator (\ref{2.14}) deviate from the atomic states, suggested in  \cite{Balykin2005}, and, as justified by our numerical simulation, construct a unique set of states, superposed in the atomic basis, which are non-orthogonal in general case. 

Indeed, the microscopic approach adds an important correction to an intuitive vision, based on the perturbation theory, and associate the quasi-particle excitation and its symmetry properties with the non-Hermitian self-energy operator (\ref{2.14}) with its dominating anti-Hermitian part but not so negligible Hermitian contribution. For the atom, considered in its local reference frame, the geometry of the object of itself has a symmetry of the $C_{2 v}$ point group with its principal axis, coincided with the radial coordinate $\rho$, directed from the $z$-axis to the atom, see Fig.~\ref{fig1}. The $C_{2 v}$ symmetry group is Abelian and has only one-dimensional representations conventionally leading to non-degenerate eigenvalues. Nevertheless we observe the double degeneracy for the five eigenvalues, which indicates a clear signature of higher local symmetry.  Take into account that the self-energy operator, thinkable as effective Hamiltonian of the valence electron, depends on its internal variables via the polarizability tensor, which components are contracted with certain weighting factors. In turn, these waiting factors are expressed by the field propagators dressed by interaction with the medium. Surprisingly but that makes this operator invariant to any internal rotation around the axis orthogonal to the $(z,\rho)$-plane sketched by the atom and waveguide. As a consequence it obeys the symmetry of point group $D_{\infty h}$ in respect to arbitrary rotation around this axis. 
 
The manifestation of higher symmetry for the effective Hamiltonian than for the physical object results from equivalence and absence of correlations in the emission process of two counter-rotating polarization modes, defined in a local frame originated with the atom's location. Then in a Cartesian basis the slope of the transition dipole moment in the $(z,\rho)$-plane only imbalances the directional total emission into the outer space and waveguide between $z>0$ and $z<0$ but does not change the rate of emission. In our calculations, reproduced in Fig.~\ref{fig3}, we have specified the eigenstates by the modulus of the angular momentum projection onto the azimuthal direction $|M_{\phi}|$ of the cylindrical coordinate system (see Fig.~\ref{fig1}), running between $0$ and $5$. 
 
However it is noteworthy to point out that the orthogonality of the eigenstates is primary provided by the dominating anti-Hermition part of the effective Hamiltonian and is slightly violated by the small contribution of its Hermitian part. That adds an admixture of extra basis states with $|M_{\phi}|\neq n$ to the each $n$-state. Nevertheless it does not break the symmetry of the self-energy operator for itself and, in accordance with the Schur's lemma, leaves the eigenvalues (\ref{3.2}) degenerate in each subspace of the group irreducible representations. Actually the non-orthogonality is meaningful for far distant separations from the nanostructure, where the respective deviations from the atomic energy become negligibly small and the system approaches to an undisturbed spherically symmetric state of the free atom.

By concluding this section let us make the following remark. As we see from the considered example the expected Purcell-type enhancement of light emission into a nanoscale waveguide of cylindrical shape reveals a quite weak effect. For both the transition schemes the coupling of an individual atomic emitter with a silica nanofiber enhances the decay rate within 50\% at maximum. In experiments the enhancement is usually certified by a factor $\beta$, which is given by empirically normalized part of the light emitted directly into the guided mode. This quantity is normally within a few percents of the total emission \cite{Nieddu2016}, since the atom is typically trapped quite far from the fiber surface, that is confirmed by our estimates in Fig.~\ref{fig2},\ref{fig3}. Similar effect, and with the same order in its magnitude, had been earlier observed for a rubidium atom experiencing the radiative type interaction with evanescent field of the light beam reflected from a plane dielectric surface, see \cite {Spreeuw2004}. Significant enforcement of the radiative coupling would be attained once we involve many atoms periodically ordered and interacting cooperatively via the guided field, see \cite{Pivovarov2020,Pivovarov2021}. Alternatively, as suggested in \cite{Kimble2014,Laurat2022b}, one could slow down the light propagation and then enhance the emission into the guided mode if a sub-wavelength waveguide was designed as a one-dimensional photonic crystal. We examine such option in the next section.

 \subsection{A comb-type asymmetric waveguide}
\begin{figure}[tp]
\includegraphics[width=8cm]{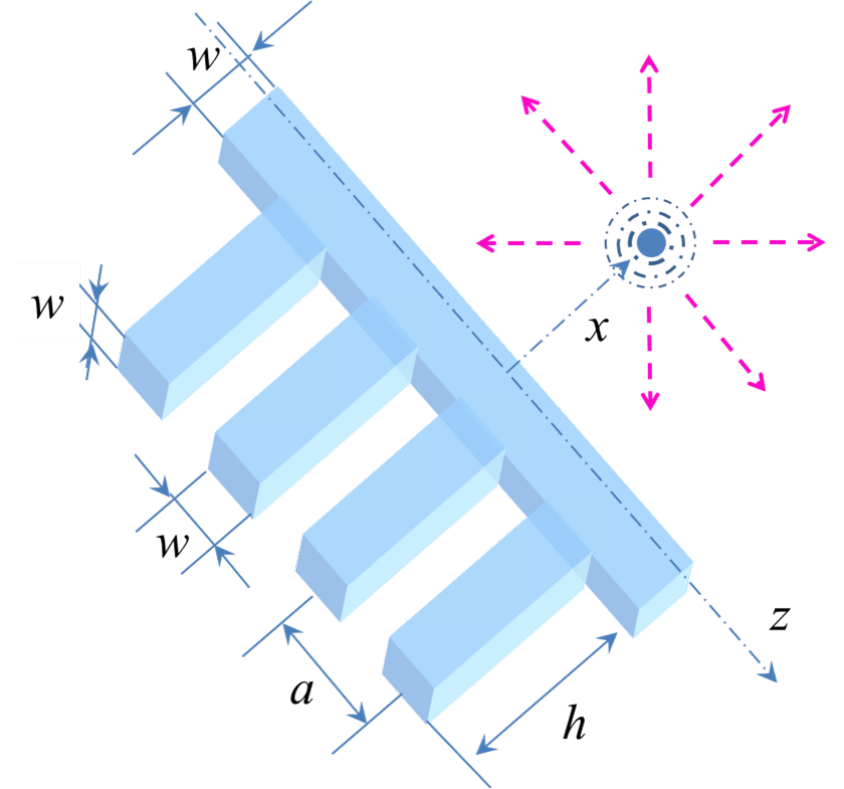}%
\caption{The single atom placed by optical tweezers (not shown) near a comb-type spatially modulated waveguide. The light emission by the atoms into the guided mode is enhanced when its frequency is tuned near the edge of the Brillouin zone of PCW and is sensitive to displacement of the atom along $z$-direction.}
\label{fig4}%
\end{figure}

\noindent In our second example we address to a specifically shaped asymmetric photonic crystal waveguide (PCW), which has a periodic spatial comb-type modulation, see Fig.~\ref{fig4}. In such a waveguide the dispersion law for light propagation in its guided mode would obey the periodic dependence as well and, as a consequence, any emission at frequency tuned near the edge of its Brillouin zone would be delayed and effectively enhanced.

For a sake of clarity and comparability with the previous example let us refer to the $F=5\to F_0=4$ closed transition in ${}^{133}$Cs with the vacuum wavelength $~852\, nm$. Then the main parameter of the comb-type waveguide is its period $a$, which is supposed to fit the longitudinal half-wavelength of the guided mode for the reference frequency and expected to be shorter than its vacuum value. For the material InGaP, suggested for experimental verification, with refractive index $\mathrm{n} \sim 3.31$, this half-wavelength is varied between $425\,nm$ (tinny waveguide) and $130\, nm$ (wide waveguide). For configuration, shown in Fig.~\ref{fig4}, we have scanned the following probe parameters: $130\, nm\!<a<\!425\,nm$, $h=1.5\,a$, $w=0.5\,a$. Particularly we shall verify the predictions followed from a macroscopic analysis reported in \cite{Laurat2022b} that such an asymmetric comb waveguide "would support (i) a slow mode with an unusual quartic dispersion around a zero-group-velocity point and (ii) an electric field that extends far into the air cladding for an optimal interaction with the atom".

For PCW we present only microscopic simulations, since there is a serious amplification for the calculation capabilities if following the macroscopic vision of the problem. The Maxwell equation cannot be analytically solved and a representative set of the waveguide modes could be only numerically constructed. Then applicability of the Fermi's golden rule would meet evident difficulties in numerical evaluation of a large number of transition matrix elements associated with these modes. To our knowledge at present there are no estimates of the decay rates for an atom coupled with a PCW in the literature. There are certain constraints for our microscopic approach as well, which hides a direct identification of the PCW parameters such as its zone structure, group velocity etc.. The method allows only indirect access to these characteristics via dependence of the calculated quantities on various external parameters.

\subsubsection{Emission of ${}^{133}$Cs at $F=5\to F_0=4$ of $D_2$-line}

\begin{figure}[tp]
\includegraphics[width=8cm]{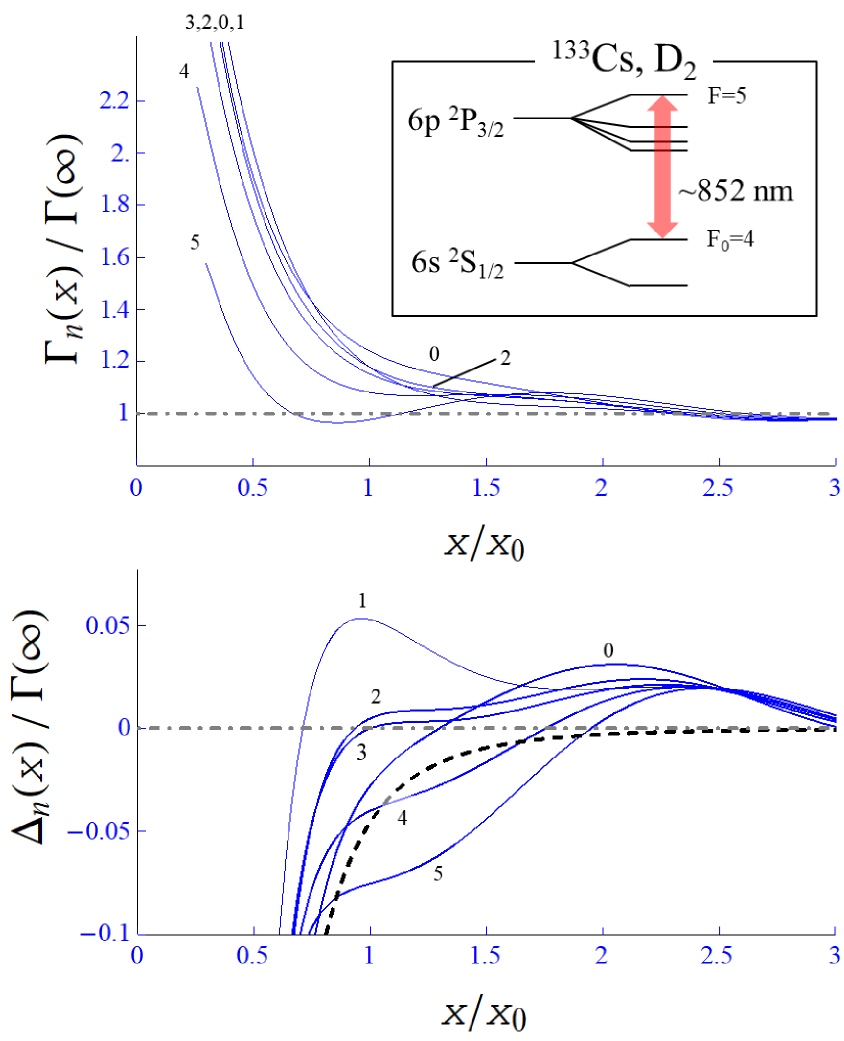}%
\caption{Same as in Fig.~\ref{fig3}, but calculated for the cesium atom coupled with the comb-type waveguide shown in Fig.~\ref{fig4}, where the atom is placed behind a comb tooth. The separation from the waveguide surface is scaled by $x_{0}=200\,nm$ similarly to Figs.~\ref{fig2} and \ref{fig3}.}
\label{fig5}%
\end{figure}

\begin{figure}[t]
\includegraphics[width=8cm]{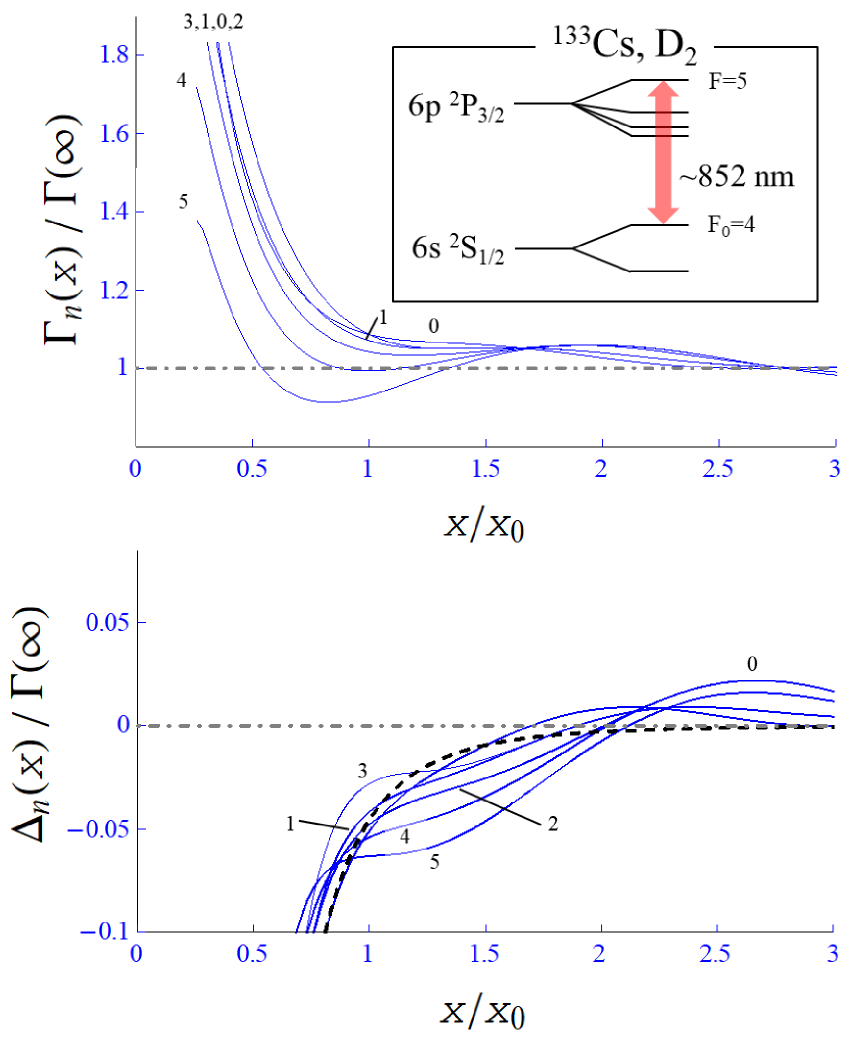}%
\caption{Same as in Fig.~\ref{fig5}, but  for the cesium atom displaced to position between the comb teeth, see Fig.~\ref{fig4}.}
\label{fig6}%
\end{figure}

\noindent In Figs.~\ref{fig5} and \ref{fig6} we show the results of our numerical simulations, presented for cesium atom, at its different locations, and for the external conditions similar to those we used in section \ref{Section_IIIA2}. From the macroscopic point of view the strongest coupling would be expected if frequency $\omega_0$ was equal to the band frequency of the Brillouin zone. We can fulfill this critical condition by varying the waveguide parameters and by searching for an optimal emission regime where the eigenvalues of the self-energy operator are maximized.

We present here the data for such an optimal configuration, which we have justified by a round of simulation cycles. With referring to Fig.~\ref{fig4}, the cesium atom is placed behind the inhomogeneity feature (comb tooth) for the data of Fig.~\ref{fig5} and behind the comb but between its teeth for the data of Fig.~\ref{fig6}. The optimal spatial period $a$ occurs a bit less than half of the vacuum wavelength and slightly different for these two cases.  We associate these deviations with a certain signature of boundary effects caused by a light reflection at the edges of the sample. Actually we deal here not with a conventional waveguide but rather with an open one-dimensional resonator. The emission enhancement is low sensitive to small variation of $a$ near its optimal value (within ten nanometers of changes) in qualitative agreement with the empirical arguments of \cite{Laurat2022b}. Our calculation accuracy has certain limitations dictated by internal convergence of the  simulation procedure in respect to extension of the sample length and density, which, in turn, is restricted by consistency in parameter variations for approaching the actual dielectric sample by its artificially constructed microscopic replica. After entire verification of the simulation protocol we have justified that the representative data correspond to the separation distances $x>100\,nm$.

The dramatic difference in the calculation results, shown in Fig.~\ref{fig3} and Figs.~\ref{fig5},\ref{fig6}, is clear seen. In the case of cylindrical waveguide and for separations from the fiber surfaces about $200\, nm$ we obtain practically negligible correction to the natural decay rate. But for the comb configuration and at the same distances there is meaningful enhancement of the emission, which is mainly expected into the guided mode. Strictly speaking our calculation scheme is incompatible with selection of the emission channel exactly into the waveguide. Nevertheless the fact, that for short separations with $x\sim 100\,nm$ we obtain the decay rate enhanced up to two times of its natural value, clear indicates that significant part of the light emerges the system primary via waveguide. 

The stronger radiative coupling with the photonic crystal is observed when the atom is placed near those inhomogeneity features where the dielectric nanostructure is locally widened and when it weakly violates the system periodic structure. As pointed out above, the periodicity slows down the group velocity of light and stimulates the emission process near the edge of the Brillouin zone if it is resonant to the atomic transition.  But we can also associate the effect of emission enhancement with optimal incorporation of the atom into a periodic structure of the photonic crystal and then into the process of Bragg diffraction, and, as a consequence, with stimulation of the directional light passage through the one-dimensional channel in general.  

Next important point is that, similarly to the case of cylindrical nanofiber, the effective Hamiltonian of the comb configured PCW reproduces a local symmetry of $D_{\infty h}$ group in respect to rotation around $y$-axis orthogonal to $(z,x)$-plane as defined in Fig.~\ref{fig4}. The eigenstates are parameterized by the modulus of the angular momentum projection $|M_{y}|$ onto the $y$-axis, which is running between $0$ and $5$. The enhanced light emission can be directionally controllable by preparation of the atom in a specific excited state. The radiative shifts, given by the real components of the complex eigenvalues, as well as the ground state van-der-Waals shift, are small but not negligible and have the same order of magnitude as in the case of cylindrical waveguide. 

\subsubsection{Emission of ${}^{87}$Rb at $F=3\to F_0=2$ of $D_2$-line and comparing remarks}

\begin{figure}[tp]
\includegraphics[width=8cm]{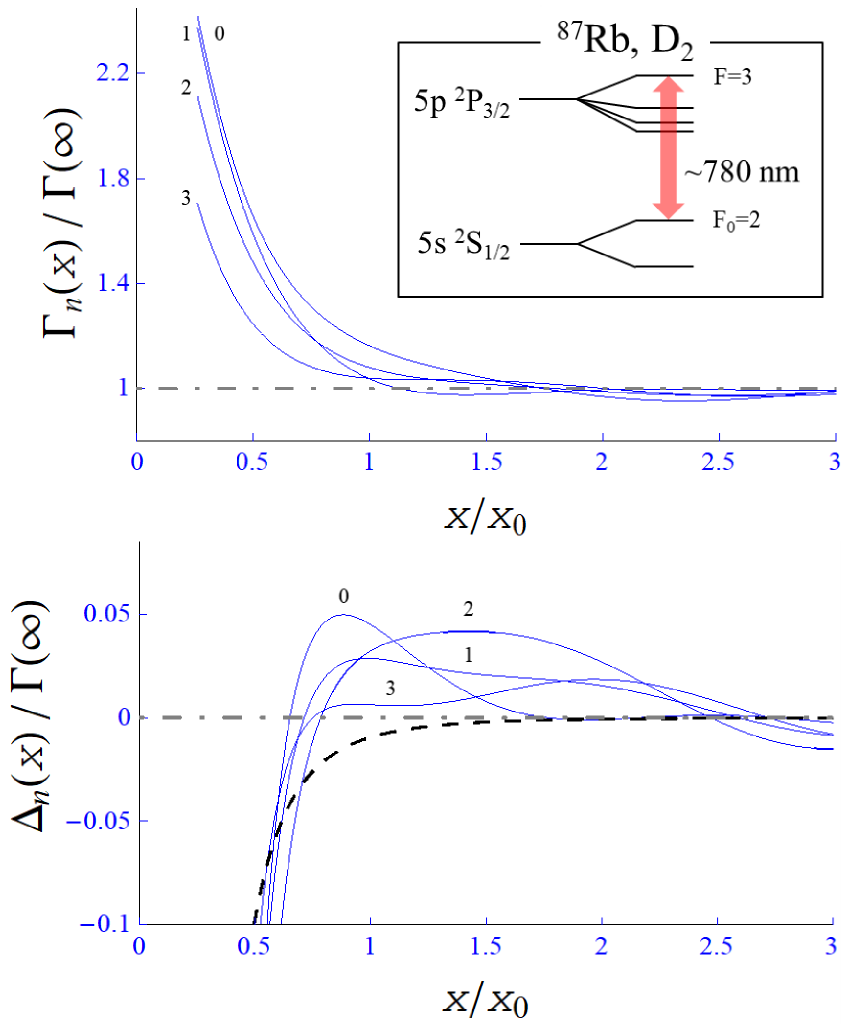}%
\caption{Same as in Fig.~\ref{fig5}, but for emission of ${}^{87}$Rb at the $F=3\to F_0=2$ transition of its $D_2$-line.}
\label{fig7}%
\end{figure}

\begin{figure}[t]
\includegraphics[width=8cm]{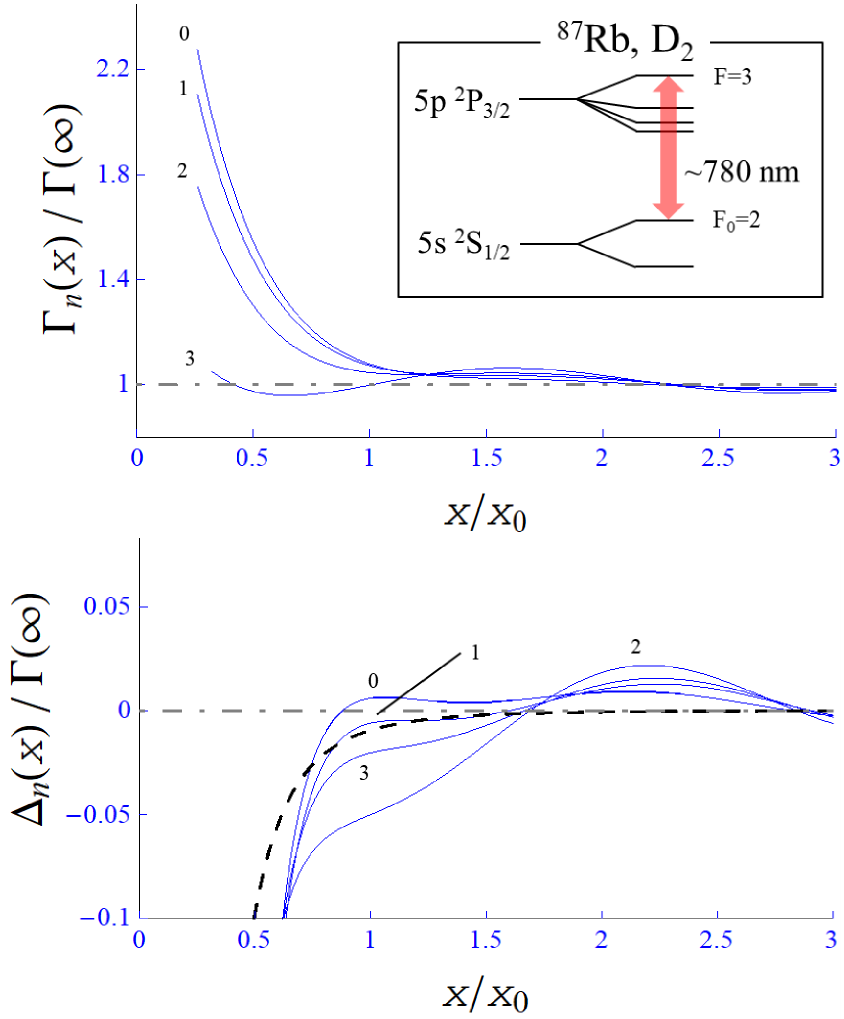}%
\caption{Same as in Fig.~\ref{fig7}, but for the rubidium atom displaced to position between the comb teeth, see Fig.~\ref{fig4}.}
\label{fig8}%
\end{figure}

\noindent For comparison in Figs.~\ref{fig7},\ref{fig8}  we show the emission rates and radiative shifts of the quasi-energy states, calculated for rubidium atom, in the same external conditions as for cesium in Figs.~\ref{fig5},\ref{fig6}. The results look qualitatively the same but give us a few additional observations. Since the rubidium has a shorter optical wavelength than cesium, the evanescent field of the guided mode is weaker in its case and, as a consequence, leads to a smaller coupling with the atom and less correction to the decay rate for the same separations from the dielectric surface. 

Next interesting point is that for both the atoms the states with $n=\max(|M_y|)=3$ (rubidium) and with $n=\max(|M_y|)=5$ (cesium) have their decay rate close to the reference values for the isolated atoms or even smaller of it at some distances. Such states can decay only onto the lower energy states with maximal projection of their spin angular momentum and the transition dipole moment lies in the $(z,\rho)$ or $(z,x)$-plane i.e. is oriented along the waveguide. Just for these states the atomic dipole has an ability for destructive interference with its image created by the dielectric sample. The effect is small but foreseen from the behaviour of all the lower curves in the upper plots of Figs.~\ref{fig3},\ref{fig6}-\ref{fig8}. The compound system (atomic dipole and nonostructure) demonstrates here a certain signature of sub-radiant regime in the emission process. From the macroscopic vision we can treat this effect as the atomic dipole together with its image have created a Dicke-type sub-radiant mode for certain separations.

For both the the atoms and for all the considered geometries the shift corrections to the quasi-energies for the atomic states reveal quite small effect. For some calculated data it is problematic to distinguish small splittings in the quasi-energy terms in the diagrams presented in Figs.~\ref{fig5}-\ref{fig8} since the observed resolution for these data is within our simulation uncertainty.  Nevertheless for rather short separations from the dielectric surface, i.e. around $100\,nm$ or less, our estimates predict a relatively strong static attraction via van-der-Waals forces for the ground as well as for the excited states.

\section{Discussion}\nopagebreak
\noindent In this paper we have presented a theoretical approach for microscopic modelling of the radiative coupling of a single atom with a dielectric nanostructure of arbitrary shape, with a main focus on nanoscale photonic crystal waveguides (PCW). The developed calculation scheme is applicable when the atom has a closed optical transition with arbitrary orbital and spin angular momenta. As we believe it can be further generalized and adjusted for more complicated objects consisting of several atoms trapped near nanostructures by optical tweezers. By the presented numerical simulations we were aiming to evaluate the potential of such physical systems for future architecture of reliable quantum interface and quantum information processing with single photons and atoms.

The excitation spectrum of the atom is described in terms of the self-energy operator responsible for the correction of the radiative dressing of the atom's excited state in the presence of a nanostructure. This operator can be also treated as an effective Hamiltonian for such a prepared atomic quasi-particle and intrinsically has a non-Hermition nature. In the considered examples of nanoscale waveguides we have obtained that the radiative interaction possesses the local symmetry of point group $D_{\infty h}$ with its principal axis directed orthogonal to the plane sketched by the waveguide and atom. In turn this leads to the double degeneracy of those quasi-particle states, which are correlated with the atomic states having non-zero projection of the total angular momentum on the symmetry axis.

The obtained results have confirmed existing expectations that PCW with asymmetric transverse profile have stronger radiative coupling between the atom and waveguide when for axially symmetric nanostructure. For the distances between the atom and the dielectric surface of about $100\,nm$ we predict the light be emitted primarily to the waveguide. That indicates a signature of strong resonance coupling between atom(s) and light propagating through the PCW in the guided mode tuned near the edge of its Brillouin zone. This in turn justifies as realistic the scenario of a signal light storage for single or few photons mapped onto a small atomic array consisting of a few atoms. This can be done by common means of either Raman or electromagnetically induced transparency protocols. If a signal photon was effectively transferred in the guided mode then we could expect, as a realistic option, to map it even onto the spin state of a single atom by the Raman storage protocol.

For the specific atomic states and at certain separations from the dielectric surface we have obtained a signature of a Dicke-type sub-radiant effect suppressing the emission rate.  The observed sub-radiant states asymptotically correlate with the states of free atom, which have maximal projection of the angular momentum onto quantization direction orthogonal to the plane sketched by the waveguide and atom. Just for such states the transition dipole moment and the waveguide lie in one plane and there is a possibility for destructive interference for the atomic dipole with its image created by the dielectric sample. The effect is obtained for both considered examples of cylindrical waveguide and PCW.

If specific sub-radiant states are excluded then in general, for a transition dipole, having component oriented orthogonal to the waveguide, we observe strong coupling and emission enhancement mainly in the guided mode. In particular, such emission channels could be activated by excitation of an atom initially occupying the upper level of the so called "clock transition" in the Zeeman manifold of its ground state, i.e. for the state having zero projection of the spin angular momentum. In this regard let us draw attention to recent experiments \cite{CGCGSKL2016,SBKISMPA2016}, where a strong cooperative interaction of alkali-metal atoms, periodically arrayed along a nanoscale waveguide, was observed via the mechanism of one-dimensional Bragg scattering at the level of only a few percents of emission enhancement per atom. In the case of PCW we expect stronger resonance scattering for a small number of atoms and even for a single atom. As we believe this suggests a promising option for organization of a quantum bus for a system of qubits adjusting the clock transition for quantum data processing.

Then, as a potentially important implication, the expected strong one-dimensional resonance scattering can provide the entanglement of atoms integrated into a quantum register. In the example of alkali-metal atoms, any detection event of elastic resonance scattering, observed in the one-dimensional channel, changes the phase of the atomic state by $-1$ if at least one or several atoms occupy the upper hyperfine sublevel and leaves the phase unchanged otherwise. So by the elastic resonance scattering of an auxiliary photon it makes it possible to create a $C_z$-type entanglement of the ground state atomic spins, which could be even more effective than the conventional protocol based on Rydberg blockade. The suggested entanglement protocol has certain robustness to weak disturbances in the atom positions.\\

\acknowledgments

\noindent
We thank Charles Sukenik for many productive discussions of experiments on light scattering from cold atoms. This work was supported by the Russian Science Foundation under Grants No. 18-72-10039 and No. 23-72-10012. D.V.K. and L.V.G. acknowledge support from Rosatom in framework of the Roadmap for Quantum computing (Contract No. 868-1.3-15/15-2021 dated October 5, 2021 and Contract No. P2154 dated November 24, 2021). A.M. acknowledges project no. 23-06308S of the Czech Science Foundation. L.V.G. and N.A.M. acknowledge support from the Foundation for the Advancement of Theoretical Physics and Mathematics “BASIS” under Grant No. 23-1-2-37-1.

\appendix
\section{The dielectric medium approached by a disordered ensemble of two-level atoms}\label{Appendix_A}
\begin{figure}[pt]
\includegraphics[width=4cm]{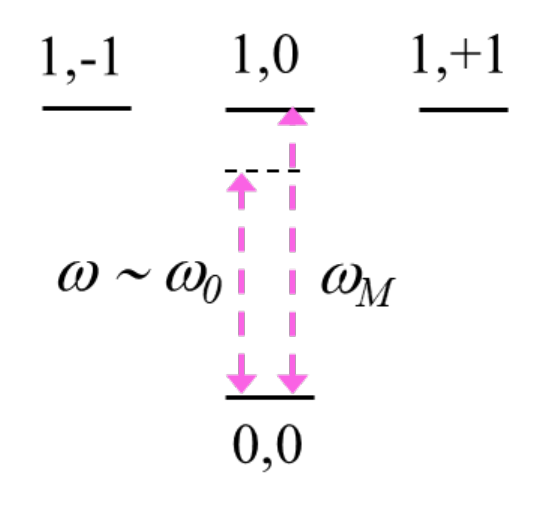}%
\caption{Energy structure of a $V$-type atom constructing the artificial medium approaching the dielectric sample. The Zeeman states are specified by the angular momentum and its projection.}
\label{fig9}%
\end{figure}

\noindent Let us consider an artificially constructed homogeneous medium consisting of two level atoms having the energy structure shown in Fig.~\ref{fig9}. The medium atom has its single ${}^{1}S_0$ ground state ($|g\rangle=|0,0\rangle$) and degenerate ${}^{1}P_1$ excited state ($|e\rangle=|1,-1\rangle;|1,0\rangle;|1,+1\rangle$), both parameterized by the electron orbital momentum and its projection. Its resonance frequency $\omega_M\equiv\omega_{eg}$ is upshifted from the resonance frequency of the reference atom $\omega_0$. The atoms are randomly distributed in space and fill the sample volume with a density $n_0$, which is high and $n_0\lambdabar_M^3>1$, where $\lambdabar_M=c/\omega_M$

As shown in \cite{SKKH2009} such a medium has its dielectric permittivity $\epsilon=\epsilon(\omega)$ given by a root of the following equation
\begin{equation}
\epsilon(\omega)=\frac{\displaystyle{1-\frac{8\pi n_0}{3\hbar}\frac{f_0^2}{\omega-\omega_M+i\sqrt{\epsilon(\omega)}\Gamma_e/2}}}%
{\displaystyle{1+\frac{4\pi n_0}{3\hbar}\frac{f_0^2}{\omega-\omega_M+i\sqrt{\epsilon(\omega)}\Gamma_e/2}}},%
\label{a.1}
\end{equation}
where $f_0$ is modulus of the transition dipole moment, and $\Gamma_e$ is spontaneous decay rate of the atomic upper state
\begin{equation}
\Gamma_e=\frac{4\omega_M^3}{3\hbar c^3}f_0^2.%
\label{a.2}
\end{equation}
Eq.~(\ref{a.1}) reveals a cubic equation, which relevant root should be selected by correct asymptotic behavior at low and high frequencies and obey the Kramers-Kronig causal relations. In particular, if the frequency argument is taken near the resonance frequency of the reference atom $\omega\sim\omega_0$ and the frequency offset is large enough, i. e. $\delta_M=\omega_M-\omega_0\gg\Gamma_e$, we arrive at the following asymptote for the real part of dielectric permittivity in a far off resonant domain
\begin{equation}
\epsilon'(\omega)\to 1-\frac{4\pi n_0}{\hbar}\frac{f_0^2}{\omega-\omega_M},%
\label{a.3}%
\end{equation}
and the complex function $\epsilon(\omega)=\epsilon'(\omega)+i\epsilon''(\omega)$ will have negligible imaginary part $\epsilon''(\omega)\ll 1$. In such an asymptotic spectral region the function (\ref{a.1}) is approximately constant and, being fixed by condition $\epsilon'(\omega_0)=\varepsilon$, will fit the permittivity $\varepsilon$ of the actual dielectric sample in a rather broad spectral domain. There are free parameters, namely, the atomic density $n_0$, the transition dipole moment $f_0$, and the frequency offset $\delta_M$, which could be independently varied to optimally fulfill the required condition. As verified by our numerical simulations for the dense samples with $n_0\lambdabar_M^3\gg 1$ and at fixed $\epsilon'(\omega_0)$ the radiative coupling of the reference atom with the medium converges to a certain level further insensitive to variation of the above parameters.

\begin{figure}[tp]
\includegraphics[width=6cm]{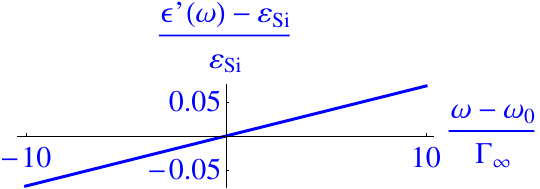}%
\caption{Fit of the spectral permittivity to the dielectric constant of silica $\varepsilon_{\mathrm{Si}}$, scanned near vicinity of the $F=0\to F_0=1$ ($D_2$-line) transition in ${}^{87}$Rb. The medium parameters are $n_0\lambdabar_M^3=20$ and $\delta_M=233\,\Gamma_{\infty}$.}
\label{fig10}%
\end{figure}

As a relevant example in Fig.~\ref{fig10} we show a typical fit for the real part of the spectral permittivity of such an artificial medium to the dielectric constant of silica ($\varepsilon_{\mathrm{Si}}\sim 1.45^2$) near the resonance frequency of the reference atom. Here we have tested the $F=0\to F_0=1$ decay channel in ${}^{87}$Rb and scaled the spectral range by the decay rate of the reference atom $\Gamma_{\infty}$. The latter is defined as an asymptotic far distant limit for the self-energy operator calculated in the main text $\Gamma_{\infty}\equiv\Gamma(\infty)$. In our estimates we set $\Gamma_e=\Gamma_{\infty}$, i.e. parameterized the artificial medium (replica of silica) by the same reduced transition dipole moment as for the reference atom. As we can see from the graph, for the parameters $n_0\lambdabar_M^3=20$ and $\delta_M=233\,\Gamma_{\infty}$, in a quite broad spectral domain near vicinity of the reference frequency the dielectric constant of silica is reliably reproducible.

\begin{figure}[tp]
\includegraphics[width=8cm]{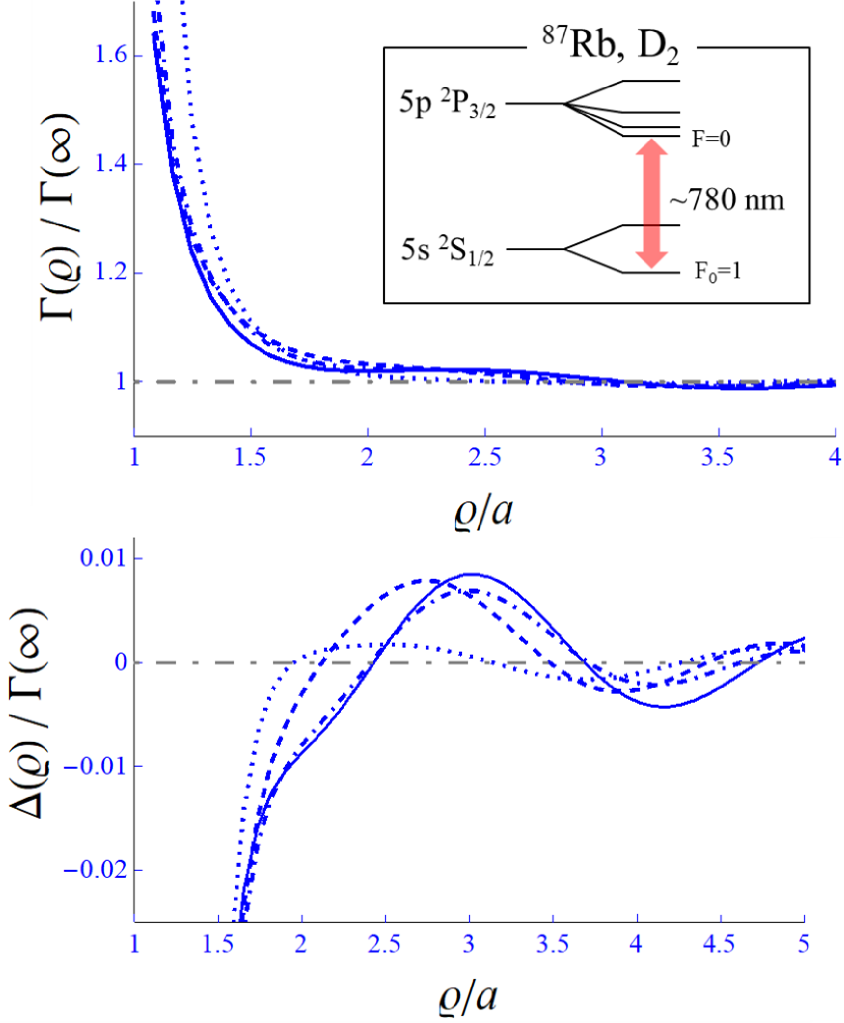}%
\caption{The decay rate and radiative shifts calculated for different densities: $n_0\lambdabar_M^3=5$ (dotted) $=10$ (dashed) $=15$ (dash-dotted) and $=20$ (solid) and with the frequency offsets respectively fitted for each density, see text.}
\label{fig11}%
\end{figure}%

\begin{figure}[tp]
\includegraphics[width=8cm]{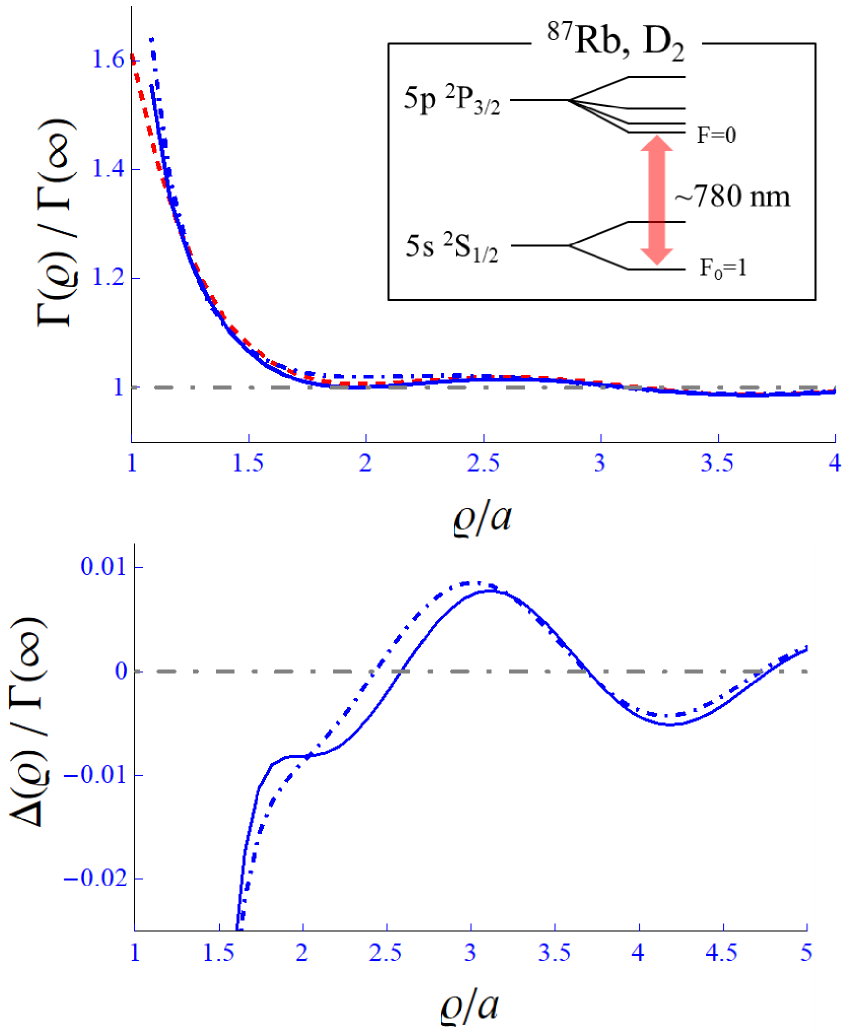}%
\caption{The decay rate and radiative shifts calculated for the ordered (solid) and disordered (dash-dotted) configurations at the density $n_0\lambdabar_M^3=20$ and for the frequency offset $\delta_M=233\,\Gamma_{\infty}$ For comparison the dashed red curve shows decay rate calculated with the Fermi's golden rule in \cite{PSGPCLK2018}.}
\label{fig12}%
\end{figure}%

The internal convergence of the method is clarified in Fig.~\ref{fig11}, where we show the results of calculations done for the different densities $n_0\lambdabar_M^3=5,10,15,20$ and respective frequency offsets $\delta_M/\Gamma_{\infty}=58, 117, 175,233$. The waveguide length was set as $4\lambda_M\sim 4\ast 780\,nm$ for all the configurations. Each curve corresponds to disordered random distribution of atoms and suggests a configuration averaging over twenty different realizations. Conceptually the dependencies are plotted in a such way that for each pair of the chosen parameters $n_0\lambdabar_M^3$ and $\delta_M$ they provide the same mean value for the dielectric constant $\varepsilon_{\mathrm{Si}}$ in vicinity of the transition frequency, see Fig.~\ref{fig10}. The dependencies for highest densities become unresolved within the graph scale.

In Fig.~\ref{fig12} we compare two different options for distributions of atoms inside the medium with $n_0\lambdabar_M^3=20$ and $\delta_M=233\,\Gamma_{\infty}$. The dash-dotted curve reproduce the calculations for the disordered medium and with the configuration averaging. The solid curve corresponds to an ordered distribution with fixed interatomic separations. The dashed red curve shows the macroscopic evaluation of the decay rate with the Fermi's golden rule. A good coincidence between the calculations done by microscopic and macroscopic approaches was already pointed in the main text. Here we have verified (and Fig.~\ref{fig12} gives an example) that the averaged parameters, calculated for the disordered sample, coincide with the limit of the ordered configuration. The latter was used in our practical calculations for all the geometries considered in the paper.

However it is necessary to point out that the variation of the external parameters cannot be infinitely ranged.  The dipole long wavelength approximation becomes inconsistent for highly dense medium where the chemical forces are evidently important. There is important constraint with the construction of the artificial medium, where the resonance frequency is assumed to be quite close to the atomic frequency in optical scale. But such condition would be impossible to fulfill if we wish to cover a broader frequency domain, than scaled by $\Gamma_{\infty}$, i.e. to apply the model beyond the example of closed transition. See also the clarifying comments after (\ref{2.28}).

\section{Van-der-Waals shift of the ground state energy}\label{Appendix_B}

\noindent The original vacuum QED diagram (\ref{2.17}) contributes not only to the excited state but to the ground state of the reference atom as well, such that it describes a light shift, induced by the vacuum fluctuations of the field, as expressed by the first line in (\ref{2.19}). Once the atom, being in the ground state, gets close to the dielectric sample it experiences a near field static interaction and the field fluctuations modified by the medium. Both the processes can be incorporated within the following extension of the original vacuum diagram
\begin{equation}
\scalebox{0.5}{\includegraphics*{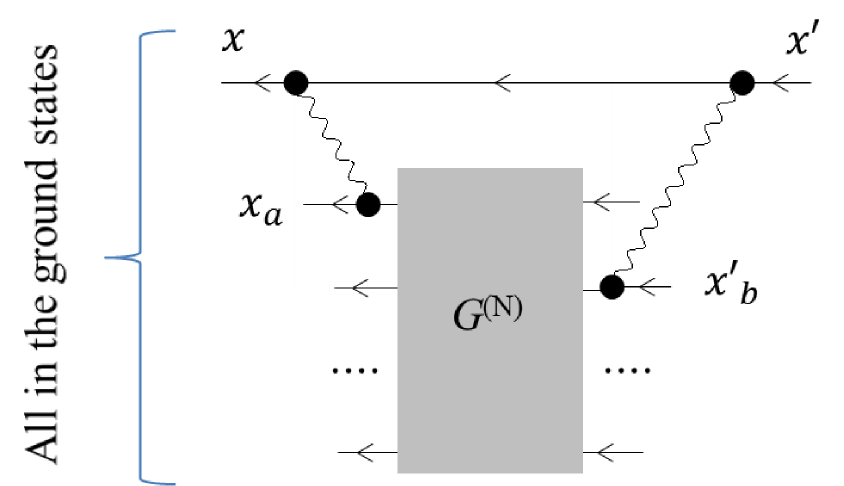}}%
\label{b.1}%
\end{equation}
This diagram describes the coupling of the reference atom with the medium in a first non-vanishing order of the perturbation theory. The shaded block symbolically visualizes the resolvent operator of the medium, consisting of $N$ atoms, with involving all internal interactions. All the inward and outward arrowed lines indicate the individual propagators associated with a stable collective ground state of the entire system.

Unlike the dynamics of a single excitation, shared by the reference atom and the medium, here we cannot follow the problem microscopically. Firstly, we deal here with far off-resonant interaction such that many intermediate energy states are involved in the virtual coupling in the diagram (\ref{b.1}). As a consequence we fail to substitute the real medium by its artificial replica constructed by ensemble of two-level atoms with simply fitting the dielectric permittivity in a vicinity of the reference transition frequency. Secondly, for a condensed medium, where atoms are confined by chemical bonds, it is certainly insufficient to be constrained by the long wavelength approximation for the internal interactions. The internal system Hamiltonian is even unknown and the medium normally yields only thermodynamically simplified macroscopic description.  Thirdly, the diagram (\ref{b.1}) is not originated with a dynamically developing process, as we have implied for an excited state. In the latter case the virtual coupling is primary initiated between singly excited states belonging to a finite segment of the system Hilbert subspace such that the excitation transport could be structured in the cooperative Dyson equation (\ref{2.15}). But unlike (\ref{2.15}) the partial contributions to (\ref{b.1}) virtually expand over infinite set of intermediate states with filling the unlimited segment in the system Hilbert subspace. In fact, (\ref{b.1}) reveals a precursor of the self-energy for the retarded-type single particle Green's function of the reference atom in its ground state, which could be approximately constructed by means of the non-equilibrium macroscopic diagram technique \cite{LfPtIX,LfPtX}.

Nevertheless, if avoiding the problem with retardation we can attempt to correct the ground state energy and constrain by the static van-der-Waals interaction only. Let us consider as a starting point, which we will further improve, the dielectric as a configuration of independent dipoles and simplify (\ref{b.1}) by only pair static-type coupling with $b=a$. Then we can approximate the interaction potential for the reference atom occupying a particular Zeeman sublevel in its ground state $|m\rangle$ by the following sum
\begin{equation}
U_{m,\underbrace{g\ldots g}_N}(\mathbf{r})=-\frac{2}{3}\sum_{a=1}^{N}\frac{1}{|\mathbf{r}-\mathbf{r}_a|^{6}}\sum_{n,e}\frac{|\mathbf{d}_{nm}|^2|\,\mathbf{f}^{(a)}_{eg}|^2}{\hbar(\omega_{nm}+\omega_{eg})}%
\label{b.2}%
\end{equation}
where the transition frequencies $\omega_{nm}$ and $\omega_{eg}$ are taken for independent atoms, and we have enumerated the medium dipoles by index $a$ running from $1$ to $N$. Here, as in the main text, we consider a single electron reference atom existing in its ${}^{2}S_{1/2}$ ground state with weak spin-orbital coupling in its excited states (negligible in the denominator of (\ref{b.2})) such that the invariant sum over intermediate states $|n\rangle$ eliminates any spin dependencies in the transition matrix elements and makes (\ref{b.2}) independent on $|m\rangle$. But the potential of itself depends on a particular microscopic configuration of the medium dipoles, which we have clarified in its subscribed index.

As an alternative macroscopic description, let us consider a homogeneous dielectric medium bounded by a flat surface and coupled with the reference atomic dipole, located at distance $z$ from the surface. Then, being in the state $|m\rangle$, the dipole experiences the static attraction from its image with the following force potential
\begin{equation}
U_m(z)=-\frac{1}{12\,z^3}\frac{\varepsilon -1}{\varepsilon+1}\sum_{n}|\mathbf{d}_{nm}|^2%
\label{b.3}%
\end{equation}
This attraction potential reveals a macroscopic counterpart of (\ref{b.2}) but with the involved internal interactions, among the medium atoms, thermally averaged over the random medium configurations. The medium response is converted into solution of the macroscopic Maxwell equation with dielectric constant $\varepsilon$ and considered for the specific plane geometry.

We can attempt to compromise (\ref{b.2}) with (\ref{b.3}) if $\omega_{nm}\ll\omega_{eg}$ for the meaningful transitions. That lets us transform the sum (\ref{b.2}) to a mesoscopically smoothed integral form
\begin{eqnarray}
\sum_{a=1}^{N}\frac{2}{3}\sum_{e}\frac{|\mathbf{f}^{(a)}_{eg}|^2}{\hbar\omega_{eg}}\ldots&\equiv&\sum_{a=1}^{N}\alpha^{(a)}\ldots\to n_0\bar{\alpha}\int\limits_{{\cal V}} d^{3}r'\ldots%
\nonumber\\%
&\approx&\frac{3}{4\pi}\frac{\varepsilon -1}{\varepsilon+2}\int\limits_{{\cal V}} d^{3}r'\ldots%
\label{b.4}%
\end{eqnarray}
with the spatial integral evaluated over the sample volume ${\cal V}$. In the last transformation, as a crucial step and extending assumption, we have included in our model the self-consistent coupling of the medium dipoles, expressed by the Lorentz-Lorenz relation between the mean polarizability $\bar{\alpha}$, averaged over mesoscopic volume and scaled by atomic density $n_0$, and dielectric constant $\varepsilon$. This last transformation is only approximately valid and could be applicable inside an infinite and homogeneous dielectric medium. In the considered case for the medium dipoles distributed near the bounding surface, which play primary role in interaction with the reference dipole, their internal coupling is overestimated by (\ref{b.4}) and expected to be weaker. But we shall apply (\ref{b.4}) as an empirical extrapolation and use it as an upper bound for the ground state static interaction beyond the low density decoding of the diagram (\ref{b.1}).

Finally we can suggest the following empirical estimate for the electrostatic shift of the ground state of the reference atom interacting via van-der Waals forces with an arbitrary shaped dielectric sample
\begin{equation}
\Delta_m(\mathbf{r})=-\frac{3}{4\pi\hbar}\frac{\varepsilon -1}{\varepsilon+2}\sum_{n}|\mathbf{d}_{nm}|^2\int\limits_{{\cal V}} d^{3}r'\frac{1}{|\mathbf{r}-\mathbf{r}'|^{6}}%
\label{b.5}%
\end{equation}
which for an infinite dielectric medium, bounded by a flat surface, coincides with (\ref{b.3}) only in assumption of $\varepsilon\gtrsim 1$. Nevertheless the deviation of the outer factor as function of $\varepsilon$ is not so critical and we believe in correctness of our estimate for arbitrary shaped nanostructures at least as its upper bound. Finally, within the made approximations the ground state shift is expected to be independent on specification of the atomic spin state, such that $\sum\limits_{n}|\mathbf{d}_{nm}|^2=\langle m|\mathbf{d}^2|m\rangle=e^2\langle r_e^2\rangle$, where $e$ is electronic charge and $\langle r_e^2\rangle$ is the variance of its radial position in the atom.

For further and more specific details concerning  van-der-Waals and Casimir forces we refer the reader to special literature \cite{LfPtIX,Sukenik1992,Sukenik1993,Ducloy1995,Lamoreaux2005}.

\smallskip


\sloppy

\bibliography{references}

\end{document}